\documentclass[a4paper,notitlepage,11pt]{article} 
\usepackage{amsfonts}
\usepackage{amssymb}
\usepackage{amsmath}
\usepackage{bm}
\usepackage{graphicx}
\usepackage{multirow}
\usepackage{color}
\usepackage{graphics}
\usepackage{cite}
\usepackage{enumitem}

\def\one{{\hbox{1\kern-.8mm l}}}

\newcommand{\beq}{\begin{equation}}
\newcommand{\eeq}{\end{equation}}
\def\be#1\ee{\begin{align}#1\end{align}}
\newcommand{\ov } {\over }

\def\vk{{\vec k}}
\def\ein{{\eta_{\text{in}}}}
\newcommand{\ri}[1]{\mbox{\text{{\tiny{\rm #1}}}}}
\newcommand{\rii}[2]{\mbox{\text{{\tiny{\rm #1}}}\hspace{-1.0pt}\text{{\tiny{\rm #2}}}}}
\newcommand{\riii}[3]{\mbox{\text{{\tiny{\rm #1}}}\hspace{-1.0pt}\text{{\tiny{\rm #2}}}\hspace{-1.0pt}\text{{\tiny{\rm #3}}}}}

\begin{document}
\begin{titlepage}
\bigskip
\begin{flushright}
\end{flushright}
\begin{center}
\vskip 2cm
\Large{Enhanced CMBR non-Gaussianities from Lorentz violation}
\end{center}
\vskip 1cm
\begin{center}
\large{Diego Chialva \\ 
      {\it Universit\'e de Mons, Service de M\'ecanique et gravitation, Place
       du Parc 20, 7000 Mons, Belgium} \\
        \tt{diego.chialva@umons.ac.be}}

\end{center}
\date{}

\pagestyle{plain}

\begin{abstract}

We study the effects of Lorentz symmetry violation
on the scalar CMBR bispectrum. 
We deal with dispersion relations modified by higher
derivative terms in a Lorentz 
breaking effective action and solve the equations via approximation techniques,
in particular the WKB method. We quantify the
degree of approximation in the computation of the bispectrum and show
how the absolute and relative errors 
can be made small at will, making the results robust.  

Our quantitative results show that
there can be enhancements in the bispectrum for specific configurations
in momentum space, when
the modified dispersion relations violate the adiabatic condition for
a short period of time in the early 
Universe. The kind of configurations that are enhanced and
the pattern of oscillations in wavenumbers that generically appear
in the bispectrum strictly depend on the form of the modified
dispersion 
relation, and therefore on the pattern of Lorentz violation. These
effects are found to be 
distinct from those that appear when modelling very high-energy
(transplanckian) physics via modified boundary conditions (modified
vacuum). In fact, under certain conditions, the enhancements are even
stronger, and possibly open a door to the experimental study of 
Lorentz violation through these phenomena.

After providing the general formulas for the bispectrum in
the presence of Lorentz violation and modified dispersion relations,
we also discuss briefly a specific example based on a healthy
modification of the Corley-Jacobson dispersion relation with negative
coefficient, and plot the shape of the
bispectrum in that case.

\end{abstract}

\end{titlepage}

\tableofcontents

\setcounter{section}{0}

\section{Introduction}

The non-Gaussian features of the anisotropies of the Cosmic Microwave
Background Radiation (CMBR) have received increasing attention in
recent times, in connection with the upcoming release of the data of
experiments like Planck \cite{Komatsu:2009kd}. The primordial scalar
non-Gaussianity occurs when the  
three- and higher-point functions of the comoving 
curvature perturbation $\zeta$ are non-zero, and are therefore a
manifestation of 
the interactions at the time of inflation \cite{Maldacena:2002vr}. However, they 
could carry signatures of physics at much higher energy, 
in particular of the so called transplanckian physics
\cite{Holman:2007na, Meerburg:2009fi}. 

In this work we are interested in the {\it bispectrum}, which is
obtained from the three-point 
function, in single-field slow-roll
models of inflation. In the conventional scenario this is slow-roll 
suppressed \cite{Maldacena:2002vr}, 
hence any non-Gaussian signature is a smoking-gun for deviation from
the standard 
picture. The effect one would generally be interested in for detection
is a peculiar shape function, that is a specific shape for the
graph of the bispectrum as a function of the external momenta, with,
in particular, 
the presence of enhancements for some of their configurations.

There have been several approaches to the study of high energy
(transplanckian) physics in cosmology: from the standard program of including
higher-derivative 
terms in the field theory action, to the modeling via 
modified uncertainty relations \cite{uncerttrans},
or via specific boundary conditions 
imposed at certain times/energy scales through boundary
actions/momenta cutoffs
\cite{boundarytrans, momcutofftrans}. In this paper, we focus on
the effects on the bispectrum due to
higher-derivative corrections yielding modified
dispersion relations for the
perturbations fields~\footnote{Our scenario differs from
  that considered in \cite{Seery:2005wm, Chen:2006nt}, where the only
  modification to the field 
equations was a change in the speed of sound (the modifications at the
level of the Lagrangian depended only on first order derivatives of
the fields). In particular, we include also
derivatives of higher order, and the effect on the field
equations is more profound.}. The dispersions differ from 
the Lorentzian one when the  physical momenta are above a certain
energy scale
$\Lambda \gg H$, where $H$ is the Hubble rate at inflation
\cite{Martin:2000xs,moddispreltrans,Lemoine:2001ar}.   

Such effects would represent a
signature of Lorentz violation at high energy and possibly open up new
opportunities to study those phenomena through the CMBR. The interest in Lorentz
violation has always been high \cite{LorentzViolReview,
  Jacobson:2000xp, Mattingly:2001yd, 
Libanov:2005yf, Jacobson:2005bg} and revived in recent times
also by
certain realization
of quantum gravity, such as the Ho\v{r}ava-Lifshitz model
\cite{Horava:2009uw}. Lorentz violation also arises in braneworld
models \cite{lorentzviolbraneworld} and modified dispersion relations
also appear in effective theory of single field inflation when the 
scalar perturbations propagate with a small sound speed 
\cite{smallsoundspeed}.

In \cite{Ashoorioon:2011eg}, the effects on the bispectrum due to modified dispersion
relations which do not violate the WKB (adiabatic) condition at early times were
studied via the specific example of the Corley-Jacobson
dispersion relation with positive quartic correction to the momentum
square term \cite{Corley:1996ar}, where the exact solution to the
field equation could be obtained. No large
enhancement factors were found, but the leading modifications to the standard
slow-roll results were strongly suppressed by the ratio 
${H^2 \ov \Lambda^2}$.  This can be
explained in terms of the very small particle creation, due to the
absence of WKB violation at early times.

In this article, we provide a more general analysis of modified dispersion relation
and their effects on the bispectrum. In particular, we consider the case
where the adiabatic condition is violated in the early Universe for
a short period of time. In this more interesting
scenario, particle 
production is more substantial. In fact, the Bogoljubov
coefficients accounting for particle production do not depend on small
ratios of energy scales such as ${H \ov \Lambda}$, but, as we will show, are
only constrained by backreaction. 
On top of that, particle creation might also lead to
enhancements factors in the bispectrum for specific configurations of the three 
external momenta, due to interference patterns. 
 
We will find that the leading modifications to the standard slow-roll 
result for the bispectrum in the presence of this kind of modified dispersion
relations exhibit two main features: 
there appear modulations (oscillations) in function of the momenta and
there can be enhancements factors for some of their configurations. We show
how their presence, their magnitude and the
enhanced configurations depend on, and can be obtained from,
the form of the modified 
dispersion relation. We also find that, under certain conditions, the
enhancements can be 
actually larger than those present within the modified vacuum
framework for transplanckian physics \cite{Holman:2007na}, in
particular when there are 
higher-derivative interaction terms. The reason
is that in the case of modified 
dispersion relations some particular cancellations in the three-point
function do not occur any more. We comment on the likelihood of
satisfying the necessary conditions for having enhancements. We also study the
backreaction issue and derive the relevant constraints descending from
it. 

The results are robust, as we will show that the errors deriving from
the use, when 
necessary, of approximations to the solution of the field equation can
be made small at will.
Our general results can be easily adapted to the theory models of interest,
obtaining from the effective action the relevant dispersion
relation. In the appendix at the end of this paper, we apply our
analysis to a particular example, which is a physical modification of
the Corley-Jacobson relation with negative coefficient, avoiding
imaginary frequencies.

The outline of the paper is as follows: we start in section
\ref{formalismnotation} by introducing the formalism for
describing Lorentz 
violation and cosmological perturbation theory, and setting the
notation. We then study the 
modified field equation for the perturbations 
and its solution in section \ref{generalsolvingequation}.
Our results regarding 
the bispectrum are presented in section \ref{bispectrumsec}: in
particular, in section \ref{cubicinteractions} we study the case of
pure cubic interactions and in section \ref{highderinteractions} the
case of cubic terms with higher derivatives.
We then
discuss the backreaction issue and find the constraints it imposes in
section \ref{backreactionsec}. We
conclude in section \ref{conclusions} with a summary and
comments. Finally, the typical Lorentz-breaking action can be read in appendix
\ref{lorbreackscalact}, and we apply our results and 
techniques to a specific example of modified dispersion relation in
section \ref{example}.

\section{Formalism and notation.}\label{formalismnotation}

The approaches to the implementation of Lorentz violation have been
various (for reviews, see \cite{LorentzViolReview}). We will prefer
the most conservative point of view, within the framework of Effective
Field Theory and preserving general covariance and
absence of fixed geometry. These latter require that Lorentz symmetry be
violated by {\em dynamical} Lorentz violating tensors, which, if we
preserve rotational symmetry in the dispersion relation, can be
reduced to (products of) vectors 
\cite{LorentzViolReview}. Preserving general covariance is clearly
appealing, because otherwise we would have to give up general
relativity. 

The standard procedure of cosmological perturbation theory can then be
followed. It begins distinguishing background
values and perturbations for the various fields. 
The cosmological background has a scale of variation much lower than
those of the Lorentz breaking corrections, and its standard
description is valid.
The distinction between background and perturbations can
be operated using different threadings and
slicings. In \cite{Maldacena:2002vr}, two gauges were used,
differing in the behaviour of the comoving curvature perturbation
$\zeta(\vec x, t)$ and the inflaton perturbation $\varphi(\vec x, t)$ 
(we indicate with $\phi(t)$ its background value). In one gauge
 \beq
  \zeta(\vec x, t) = 0, \qquad  \varphi(\vec x, t) \neq 0 \, ,
 \eeq
in the other
 \beq
  \zeta(\vec x, t) \neq 0, \qquad  \varphi(\vec x, t) = 0 .
 \eeq
The latter gauge is more convenient conceptually, as one works
directly with the physically interesting perturbation (the curvature
one, which is conserved outside the horizon), but the former one is
often better computationally.

In our case, the first gauge is by far
the most convenient, as one can take into account the higher
derivatives correction via the action of the inflaton (a simple
scalar field). The inflaton sector will therefore be our
Lorentz-breaking sector. This is especially appealing when discussing
backreaction, as we will do in section \ref{backreactionsec}, because one does
not have to compute the effective stress-energy tensor of the metric
perturbations, which is a complicate calculation. 
A covariant action with higher-derivative terms and a
dynamical Lorentz-breaking vector field can be constructed
\cite{Lemoine:2001ar, Jacobson:2000xp, Mattingly:2001yd}; we report it
in appendix \ref{lorbreackscalact}. 

The transformation
between the two gauges is a time-reparametrization $t \to t+T$, such
that\footnote{Derivatives with respect to $t$ are indicated with the
  notation $_{,t}$, those with respect to
  the conformal time $\eta$ with $'$.} 
 \beq \label{gaugechangezetavarphi}
  \zeta =  H T + {1 \ov 2} H_{,t} T^2 - 
   {1 \ov 4} \partial_i T \partial_j T a^{- 2} +
   {1 \ov 2} \partial_i \chi \partial_i T  
    + {1\ov 4} a^{- 2} \partial^{-2} \partial_i
    \partial_j \left( \partial_i T \partial_j T \right)
    - {1 \ov 2}\partial^{-2} \partial_i\partial_j 
    \left(\partial_i\chi  \partial_j T \right) 
 \eeq
where $H$ and $a$ are, respectively, the Hubble rate and the metric
scale factor, while
 \beq
  T = - {\varphi \ov \phi_{,t}} - {1 \ov 2} 
   {(\phi_{,t})_{,t}  \varphi^2 \ov (\phi_{,t})^3} + 
   {\varphi_{,t} \varphi \ov (\phi_{,t})^2} \, ,
 \qquad 
  \partial^2 \chi = {(\phi_{,t})^2 \ov H^2 }  
  {d \ov dt}\biggl( - {H \over \phi_{,t}} \varphi \biggr).
 \eeq
The higher order terms in $\varphi$ and in the slow-roll parameter
$\varepsilon = {\dot\phi^2 \ov 2H^2 M_{_\text{Planck}}^2}$ are necessary to
compensate for the time evolution of $\varphi$ on superhorizon scales
and make $\zeta$ conserved outside the horizon. For our needs, it
suffices to consider this relation 
only at leading order in perturbations and slow roll
parameters.
 
We expand the inflaton perturbation $\varphi$ as
 \beq
  \varphi(\eta, x) = \int {d^3 k \ov (2\pi)^{{3 \over 2}}}
  \varphi_{\vk}(\eta) 
     e^{i \vec{k} \cdot \vec{x}} \, ,
 \eeq
using the conformal time $\eta = \int dt a^{-1}(t)$, and quantize writing
 \beq
  \varphi_{\vk}(\eta) = f_{\vk}(\eta) \hat a_\vk^\dagger +
  f_{\vk}^*(\eta) \hat a_\vk \, ,
 \eeq
 where
$f_{\vk}(\eta), f_{\vk}^*(\eta)$ are two linearly independent
 solutions of the field equation
 \beq \label{eqofmo}
   f_\vk'' + \left(\omega(\eta,\vk)^2- {z'' \ov z} \right) f_\vk = 0
    \qquad z = {a\phi_{,t} \ov H} \, .
 \eeq
Different choices for $f_{\vk}^*(\eta), f_{\vk}(\eta)$ correspond to
different choices of vacuum for the field.
Imposing the standard commutation relation on the operators
$\hat a_{_\vk}~^{\text{\tiny $\dagger$}}, \hat a_{_\vk}$, implies a certain normalization for the
Wronskian of $f_{\vk}^*(\eta), f_{\vk}(\eta)$.
Using equation (\ref{gaugechangezetavarphi}) at leading order, the
comoving curvature perturbation is then expanded as
 \beq
  \zeta(\eta, x) = \int {d^3 k \ov (2\pi)^{{3 \over 2}}} \zeta_{\vk}(\eta)
     e^{i \vec{k} \cdot \vec{x}} \, ,
  \qquad \zeta_{\vk}(\eta) = {\varphi_{\vk}(\eta) \ov z} \, .
 \eeq

In equation (\ref{eqofmo}), $\omega(\eta,\vk)$ is the comoving
frequency as read from the  
effective action. In the
standard Lorentzian case $\omega(\eta,\vk)$ is 
equal to $k\equiv |\vk|$, but for a modified dispersion 
relation $\omega$ will have a different dependence 
on $\vk, \eta$. For isotropic backgrounds, which we will limit
ourselves to, $f_{\vk}(\eta)$ and
$\omega(\eta,\vk)$ depend only on $k$ and, therefore, we may drop
the arrow symbol in the following.

\section{Modified dispersion relations, field equations and WKB
  method.}\label{generalsolvingequation}  

We rewrite here for convenience the field equation (\ref{eqofmo}) for
the mode functions: 
 \beq \label{eqofmoUpot}
   f_\vk'' + V(\eta, \vk)^2 f_\vk = 0
   \qquad V(\eta, \vk)^2 \equiv \omega(\eta, \vk)^2- {z'' \ov z}.
 \eeq 
Solving it in the case of a modified
dispersion relation is often difficult and
approximation methods have to be employed. One of them consists in
exploiting the WKB approximation where
possible.
This approach provides a global
approximated solution to differential equations whose highest
derivative term is multiplied by a small parameter that we call $\epsilon$
\cite{BenderOrszagMathMeth}.
We will therefore
rewrite equation (\ref{eqofmoUpot}) in such a way that the
WKB method can be rigorously applied. Our approach will somehow
differ from what has been done in the past in the cosmological
literature discussing the spectrum of perturbations, and will
enable us to have a better control on the 
approximation.

It is useful to start with a brief general review of the WKB method to
make the paper 
self-contained and fix the notation; more details can
be found in \cite{BenderOrszagMathMeth}. The method applies to
equations of the form 
 \beq
  \epsilon {d^n f \ov dy^n}+ a(y){d^{n-1} f \ov dy^{n-1}}+
  b(y){d^{n-2} f \ov dy^{n-2}} + \cdots + m(y) f = 0 \, ,
 \eeq
but it can also be used for inhomogeneous equations, to approximate the
Green functions. In our case the equations are of 
second order and can be put in the form
 \beq \label{standardWKBeq}
  \epsilon^2 {d^2 f \ov dy^2} + q(y) f = 0
 \eeq 
The WKB method consists then in: 
\begin{itemize}
 \item[{\em i})] postulating an approximated solution of the form
   \beq \label{WKBexp} 
    f(y) \sim e^{{1 \ov \delta} \sum_{n\geq 0}\delta^n s_n} \, ,
   \eeq
  for a small parameter $\delta$ that will be fixed by studying the
  differential equation as we will show,  
 \item[{\em ii})]
  substituting the ansatz (\ref{WKBexp}) in the equation
  (\ref{standardWKBeq}), obtaining\footnote{From now on we indicate
    with dots the
  derivatives with respect to the variable $y$ for which the
  differential equation at hands has the form
  (\ref{standardWKBeq}). The apex $'$ will keep denoting the derivation
  with respect to $\eta$.}
   \beq \label{TotSequenceEquations} 
   {\epsilon^2 \ov \delta^2} \dot s_0^2+ {\epsilon^2 \ov \delta} 2\dot s_0 \dot s_1
   + {\epsilon^2 \ov \delta} \ddot s_0 + \epsilon^2 \, \dot s_1^2
   + \epsilon^2 \, 2 \dot s_0 \dot s_2 + \epsilon^2 \, \ddot s_1 
   + O(\epsilon^2\, \delta) +q(y) = 0 \, ,
   \eeq
 \item[{\em iii})]
  determining the value of $\delta$ through a {\em distinguishing
    limit} (see \cite{BenderOrszagMathMeth} and in the following) and solving the sequence
  of equations obtained from (\ref{TotSequenceEquations}) by comparing
  powers of $\epsilon$. 
\end{itemize}
We will soon give practical examples of these steps when presenting
the cosmological application. Before that, let us discuss the
validity conditions for the WKB approximation.
The first condition is that the series at the exponent of
(\ref{WKBexp}) is an {\em asymptotic} series uniformly for all the
interval of $x$ considered. Moreover, the validity of the procedure
for solving the differential equations requires/guarantees that the
series can be differentiated term-wise (at least up to the order of the
equations) and the derivative series are all asymptotic as well. We
then need to satisfy \cite{BenderOrszagMathMeth}
 \beq \label{WKBseries}
  \delta s_{n+1} \sim o(s_{n}) \to  \biggl|{\delta s_{n+1} \ov s_{n}}\biggr| \ll 1,
  , \qquad
  \delta \dot s_{n+1} \sim o(\dot s_{n}) \to  \biggl|{\delta \dot s_{n+1} \ov \dot s_{n}}\biggr| \ll 1, 
  \qquad \cdots \quad \forall \, n, \qquad \delta \to 0 \,  
 \eeq
uniformly for the interval of $x$ considered.

Furthermore, since the asymptotic
series is at the exponential, these conditions are not enough to
guarantee that we have a good approximation of $f(y)$: for the WKB
series truncated at the order $m$ to be a good approximation of $f(y)$
we must also require
 \beq \label{WKBapprox}
  \delta^{m} s_{m+1} \ll 1 \qquad \delta \to 0 .
 \eeq
 The relative error made when using the WKB approximation
truncated at the order $m$ is then
 \beq
  {f(y) - e^{{1 \ov \delta} \sum_{n = 0}^m \delta^n s_n} \ov f(y)}
  \sim \delta^{m} s_{m+1} \, .
 \eeq
We will also be concerned with the error done when integrating the WKB
approximation in place of the exact solution. It can be found that
 \begin{gather}
  \int f(y) - \int e^{{1 \ov \delta} \sum_{n = 0}^m \delta^n s_n}
  \sim \int e^{{1 \ov \delta} \sum_{n = 0}^m \delta^n s_n} \, \delta^{m} s_{m+1} \\
  \int f_1(y)f_2(y)f_3(y) - \int \prod_{i = 1}^3 e^{{1 \ov \delta} \sum_{n = 0}^m\delta^n s_{i, n}}
  \sim \int \prod_{i = 1}^3 e^{{1 \ov \delta} \sum_{n = 0}^m\delta^n s_{i, n}}
        \, \delta^{m} \sum_{j = 1}^3 s_{j, m+1}
 \end{gather}
which, beside being small because proportional to $\delta^{m}$, is
further suppressed when  
$e^{{1 \ov \delta} \sum_{n = 0}^m\delta^n s_{i, n}}$ is a rapidly
oscillating function, as it will be in our case. These criteria are
quantitative and tell us 
the order we need to go to for approximating the result to some
prescribed error, which we can make small at will by going to higher orders in the
approximation.

The conditions (\ref{WKBseries}), (\ref{WKBapprox}) are the {\em WKB
  condition(s)}. Given an 
equation of the form (\ref{standardWKBeq}), they generically require a
slow variation of the quantity $q(x)$ and are also called ``adiabatic''.

Let us now investigate the application of the WKB method to the
cosmological equation (\ref{eqofmoUpot}) in the case of modified
dispersion relations. We will 
show that there is in fact a natural choice of
variables such that a small parameter appears. 
We consider dispersion relations where the usual linear behaviour
receives non-negligible corrections when the momentum is larger than a
certain physical scale $\Lambda \gg H$, and therefore
the dispersion relation can be written in general  as~\footnote{In
  many cases, the dispersion 
  relation is given directly as an expansion in power
  series, for example 
$\omega(k, \eta)^2 \approx \sum_{n=o} c_n\bigl({H^2 \ov
    \Lambda^2}\bigr)^n(k^2)^{n+1}\eta^{2n}, \; c_0=1$; we
   however prefer to use the generic formula in
   (\ref{powerexp}). There are some discussions about the scale suppression of
   the higher derivative terms, see \cite{LorentzViolReview}.}
 \beq \label{powerexp}  
  \omega(\eta, k) = {\omega_{_\text{phys}}(p) \ov a(\eta)} 
    = {p \ov a(\eta)} \;  F\left(-{p \ov \Lambda}\right) 
    = k \;  F\left({H \ov \Lambda}k\eta\right) \,,
  \qquad F(x \to 0) \to 1 \, .
 \eeq

We now change variable to
$y = k\eta {H \ov \Lambda}$ 
and call\footnote{The slow-roll parameter 
$\varepsilon = -{1 \ov H^2}{d H \ov dt}$ 
should not be confused with $\epsilon = {H \ov \Lambda}$.} 
${H \ov \Lambda} = \epsilon$. The field equation then reads
 \beq \label{standardWKBcosmdispy} 
   \epsilon^2 \, \ddot f_y + \biggl(F(y)^2 - \epsilon^2 {2 \ov y^2}\biggr)f_y = 0 \, ,
 \eeq 
So far, we need not specify if we discuss sub- or superhorizon
scales. We proceed with the WKB method and write
(\ref{TotSequenceEquations}) with
 \beq \label{potxepsvar}
  q(y) = F(y)^2 - \epsilon^2 {2 \ov y^2} \, .
 \eeq
By looking at equations (\ref{TotSequenceEquations}) and
(\ref{potxepsvar}), we spot two possible distinguished 
limits for $\delta$, that is $\delta \sim \epsilon$ and $\delta \sim
1$. They arise from two different {\em dominant balance} conditions
\cite{BenderOrszagMathMeth}. In the first case, the leading equation
when considering powers of $\epsilon$ is
 \beq
  \dot s_0^2 + F(y)^2 = 0 
 \eeq
In the second case the leading equations are
 \beq
  \epsilon^2\biggl((\sum_n \dot s_n)^2 + \sum_n \ddot s_n -{2 \ov y^2}\biggr) = 0 
  \qquad F(y)^2 \ll \epsilon^2 {2 \ov y^2}
 \eeq
Here, it surfaces the distinction between superHubble and
subHubble modes, which corresponds respectively to the first and
second distinguished limit for $\delta$. In section
\ref{fieldsolution}, we will use the WKB
approximation in the intervals of time where the first limit is
appropriate, and look for a different solution in the case of
the second limit. In fact, in this latter case, the WKB conditions
(\ref{WKBseries}), (\ref{WKBapprox}) will not be 
satisfied~\footnote{\label{ImprovingsuperhorizonWKB} It might happen
  that with an appropriate change of coordinates the WKB approximation could be
improved also in the region where it would not be good when using the
conformal time variable $\eta$ or the rescaled variable $y =
\epsilon\, k\eta$, see for example what
  shown in \cite{Martin:2002vn} for the case of standard dispersion relations.
However, we will consider the general case and keep using $\eta, y$
and speaking of WKB violation 
when the WKB conditions are not met in these coordinates.}.

For the moment, let us discuss which kind of modified dispersion
relations can appear. In general, they can be divided in two different
classes, on the basis of the behaviour of the frequency 
at early times and the value of the Bogoljubov
parameters. 

The first class is represented by those dispersion
relations for which there is no
violation of the WKB condition at early times, but only at very late
times when the dispersion has the linear form
$\omega \approx k$ as in the standard case. 
A good example within this class is the Corley-Jacobson
relation with positive coefficients \cite{Corley:1996ar}, which was
studied in 
\cite{Ashoorioon:2011eg}. In this case, the WKB
approximation (well-suited for these relations) shows that
the corrections to the spectrum and bispectrum are very suppressed.

The second class of dispersion relations consists instead of those
violating the WKB (adiabatic) condition at early times. 
This scenario is more promising: first, interference terms  
will generally appear in the bispectrum due to the presence of both
positive 
and negative frequency solutions in consequence of the early WKB
violation. In this case, one could therefore expect large 
enhancement factors for some configurations because of interference
and phase cancellation. 
Second, the Bogoljubov coefficients indicating
particle creation are actually 
quite independent of any ratio of 
scales of the form ${H \ov \Lambda}$, as we will show, and are
proportional 
to the parameter signalling the WKB violation and the interval 
of time during which the WKB 
approximation is not good. The only constraint is due to 
the request of small 
backreaction \cite{moddispreltrans}.

\begin{figure}[ht!]
\centering
\includegraphics*[width=250pt, height=180pt]{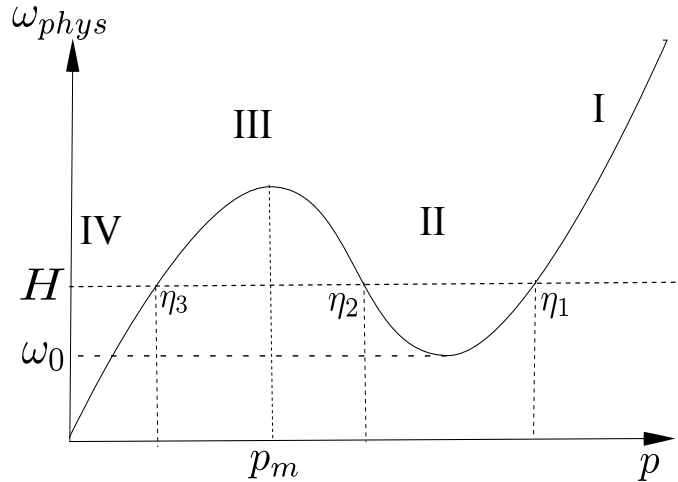}
\caption{Dispersion relation with violation of WKB at early times.}
\label{INOUTdispersion}
\end{figure}

In the following, we will study the corrections to the bispectrum
for dispersion relations with the behaviour represented in 
figure \ref{INOUTdispersion}, where the WKB condition is violated at
early times. The times $\eta_{\ri I},
\eta_{\rii II}, \eta_{\riii III}$ signalling the transition between 
violated/satisfied WKB condition are individuated by the
violation of (some of) the conditions in equations
(\ref{WKBseries}), (\ref{WKBapprox}), and are close 
to the turning points $\eta_1, \eta_2, \eta_3$ in figure 
\ref{INOUTdispersion}. They are inversely proportional to the
wavenumber $k$, being given by a condition on $y=\epsilon k \eta$,
and hence, more 
correctly, we will indicate them with the notation $\eta^{(k)}_{j}$
when necessary to avoid confusion. It is important that at very late times the WKB condition
(adiabaticity) is restored, because in that case we can unambiguously
determine a vacuum state and fix initial data. We will also briefly
comment on the case of dispersion relations with no WKB
violation at 
early times, generalizing the results found in \cite{Ashoorioon:2011eg}. 

Dispersion relations that are more complicated than the one in figure
\ref{INOUTdispersion} can be imagined, with several period of WKB
violation and a more varied behaviour, but the analysis in those
cases amounts to repeating the one we do for the case in
figure \ref{INOUTdispersion} in the various regions of WKB validity or
violation.

\subsection{Solving the field equations}\label{fieldsolution}

In the case of a dispersion relation like the one in figure
\ref{INOUTdispersion}, the time axis can be divided in four regions,
and the solution written as follows 
 \beq \label{piecewisesolution}
 f_\vk(\eta) = 
  \begin{cases}
  \varsigma_\vk \, u_1(\eta, \vk) & 
     \qquad {\rm I} : \; \eta < \eta^{(k)}_{\ri I} \\
  B_1 \, \mathcal{U}_1(\eta, \vk) + B_2 \, \mathcal{U}_2(\eta, \vk) 
    & \qquad {\rm II} : \; \eta^{(k)}_{\ri I} < \eta < \eta^{(k)}_{\rii II} \\
  \alpha_\vk \, u_1(\eta, \vk) + \beta_\vk \, u_2(\eta, \vk)
    & \qquad {\rm III} : \; \eta^{(k)}_{\rii II} < \eta < \eta^{(k)}_{\riii III} \\
  D_1 \, \mathcal{V}_1(\eta, \vk) + D_2 \,\mathcal{V}_2(\eta, \vk)
  & \qquad {\rm IV} : \; \eta^{(k)}_{\riii III} < \eta  
 \end{cases}
 \eeq
The coefficients $\alpha_\vk, \beta_\vk, B_{1,2}, D_{1,2}$ are
determined by imposing continuity for the function and its first
derivative, $\varsigma_\vk$ is determined by the choice of boundary
conditions for $\eta \to -\infty$.

We also must impose the Wronskian condition $\mathcal{W}\{f, f^*\}
= -i$ to have the standard commutation relations in the quantum theory.
If we impose it at a
certain time, the continuity of the function and its first
derivative ensures that it will be always satisfied because of Abel's
theorem and equation (\ref{eqofmoUpot}). 

The partial solutions listed in
equation (\ref{piecewisesolution}) will be presented in the next
sections; in particular, we will use the WKB method to obtain 
$u_{1, 2}$ and choose the adiabatic vacuum for the field.

\subsubsection{The WKB solution of the field equations in regions {\rm
I} and {\rm III}}\label{WKBfieldsolution} 

As we have shown, the WKB method has lead us to the equation
(\ref{TotSequenceEquations}) with 
$q(y)$ given by (\ref{potxepsvar}), which
can be solved in sequence comparing powers of $\epsilon$. In the
regions {\rm I}, {\rm III} the correct distinguished limit is $\delta
\sim \epsilon$, leading to the sequence
 \beq
 \begin{cases}
  \dot s_0^2 + F(y)^2 = 0 \\
  2 \dot s_0 \dot s_1 + \ddot s_0 = 0 \\
  \dot s_1^2 + 2 \dot s_0 \dot s_2 + \ddot s_1 -{2 \ov x^2} = 0 \\
  2 \dot s_1 \dot s_2 + 2 \dot s_0 \dot s_3 + \ddot s_2 = 0 \\
  \cdots
 \end{cases} \, .
 \eeq

To take into account the
${z'' \ov z}$ term in the field equation, we need to include up to the order
$\delta^2$ in the series expansion for the WKB solution. The Wronskian
condition then forces us to consider also the order $\delta^3$. 
In this way
we obtain the positive- and negative-frequency WKB
solutions\footnote{From the ansatz (\ref{WKBexp}) of the WKB solution
  we actually obtain 
$e^{\pm{i \ov \epsilon}\Omega(y_k)-{1\ov 2}\log(S_0) - {\epsilon^2 \ov 2}{S_2 \ov S_0}}$ 
instead than 
$e^{\pm{i \ov \epsilon}\Omega(y_k)-{1\ov 2}\log(S_0+\epsilon^2 S_2)}$, but the
  resummed form is better for the Wronskian condition. Here, recall 
  $U = S_0 + \epsilon^2 S_2$.} $u_{1, 2}$
 \beq
  u_1(y_k) =
  {e^{-{i \ov \epsilon}\Omega(y_k)} \ov \sqrt{2 \, k \, U(y_k)}} \, , 
  \qquad
  u_2(y_k) = u_1^*(y_k) =  
  {e^{+{i \ov \epsilon}\Omega(y_k)} \ov \sqrt{2 \, k \, U(y_k)}}
  \, ,
 \eeq
 \beq \label{Omegaphase}
  \Omega(y_k)=\int^{y_k} U(y'_k) dy'_k
  \, , \qquad 
  U(y_k) \equiv S_0+\epsilon^2 S_2 \equiv F + \epsilon^2 \biggl(-{\ddot F \ov 4 F^2}
  + 3{(\dot F)^{2} \ov 8 F^3} - {1 \ov F y_k^2}\biggr) \, .
 \eeq
With the above notation for $\Omega(y_k)$ we intend the primitive
of $U$ evaluated at $y_k = \epsilon k \eta$, since we absorb the
phases resulting from the lower limit
of integration in the coefficients
$\varsigma_\vk, \alpha_\vk, \beta_\vk$. The notation $y_k$
keeps track of the $k$-dependence of the definition of $y$ and will be
useful in the following.
We have also normalized the solution with a factor $\sqrt{k}^{-1}$
because in this way the Wronskian constraint  
$\mathcal{W}\{u_1, u_2\} = -i$ in the $\eta$-coordinate is
solved if $\varsigma_\vk$ is a pure phase, 
and $|\alpha_\vk|^2-|\beta_\vk|^2 = 1$. 

The WKB approximation is good when all the conditions (\ref{WKBseries}),
(\ref{WKBapprox}) are satisfied. In particular, since we truncated the
series at the order $\delta^3$, we must require, among the other conditions,
 \beq \label{WKBconditionorderfour}
  \epsilon^4 \biggl|{\dot s_4 \ov \dot s_0}\biggr| \ll 1 \, .
 \eeq
Observe that, by direct
derivation, our WKB solutions are found to
precisely satisfy the equation (using the coordinate $\eta$)
 \beq \label{WKBequation}
    u_{1, 2}'' + \bigl[V(\eta, \vk)^2 -\mathcal{Q}(\eta, \vk)\bigr]u_{1, 2} = 0  
 \eeq
where
 \beq \label{WKBparameteratorderfour} 
  \mathcal{Q}  = -\biggl(-{\omega'' \ov 4 \omega^2}
  + 3{\omega^{'2} \ov 8 \omega^3} - {1 \ov \omega y^2}\biggr)^2 
  + {\omega'' \ov 2 \omega} - 3{\omega^{'2} \ov 4 \omega^2}
  -{U'' \ov 2 U} + 3{U^{'2} \ov 4 U^2}
  \underset{\eta \to y_k}{\simeq} -\epsilon^4 2 \dot s_0 (y_k) \dot s_4(y_k) \, ,
 \eeq
therefore the condition (\ref{WKBconditionorderfour}) corresponds to
$\bigl|{\mathcal{Q} \ov V^2}\bigr| \ll 1$.
The times $\eta_{\rii II}, \eta_{\riii III}$ are then such that 
$\bigl|{\mathcal{Q} \ov V^2}\bigr|_{\eta_{\rii II, \riii III}} \not\ll 1$.

\subsubsection{The solution of the field equations in regions
  {\rm II} and {\rm IV}}\label{NonWKBfieldsolution} 

The WKB approximation is not good in the regions {\rm II}
and  {\rm IV}~\footnote{See however footnote
  \ref{ImprovingsuperhorizonWKB}.}.
In previous works, then, it has then often been taken the drastic
approximation 
that $V(\eta, \vk)^2 \approx {z^{''} \ov z}$ in those regions, so that
the solution of 
equation (\ref{eqofmoUpot}) reduced to 
a growing and decaying mode depending on the scale factor.
For the purpose of studying the bispectrum, we employ more refined
and accurate approximations.

Indeed, in region {\rm IV}, when $\eta > \eta_{\riii III}$, the dispersion relation is in the
linear regime 
 \beq \label{latelinearomega}
  \omega^2(\eta > \eta_{\riii III}, \vk) = k^2,
 \eeq 
and the solution is given in terms of the Hankel functions 
$\mathcal{V}_{1, 2} = \sqrt{-\eta} H^{(1, 2)}_\nu(-k\eta), \; \nu \sim {3 \ov 2}$. 
By asking for the continuity of the
function and its first derivative at $\eta \lesssim \eta_{\riii III}$,
we find that
 \beq \label{Dcoef}
  D_1 = {\sqrt{\pi} \ov 2} e^{i{\pi \ov 2}\nu+i{\pi \ov 4}} \alpha_\vk
  \qquad  D_2 = {\sqrt{\pi} \ov 2} e^{-i{\pi \ov 2}\nu-i{\pi \ov 4}} \beta_\vk
 \eeq
 
When instead $\eta_{\ri I} < \eta < \eta_{\rii II}$, in region {\rm II}, we indicate with 
$\mathcal{U}_{1, 2}(k, \eta)$ the solutions obtained not neglecting
the $\omega^2$ term in the potential $V^2$ in equation (\ref{eqofmoUpot}). 
Actually, we will find that some of the important features of the
bispectrum can be found without knowing the details of the
field solution, but just using the fact that small backreaction requires having 
a short interval $[\eta_{\ri I}, \eta_{\rii II}]$, as we will show in section
\ref{backreactionsec}. In this sense, those
results will be very robust, being
independent of the approximations chosen when solving the field equations.

When needed, however, a rather accurate approximation in region 
{\rm II} is obtained
observing that, given figure \ref{INOUTdispersion}, the
physical frequency has a local minimum $\omega_0$  at $\eta =
\eta_m$, and therefore 
 \beq \label{omeganearmin}
  \omega^2 = a^2 \omega_{_\text{phys}}^2 \simeq 
   {1 \ov \eta^2 H^2}\biggl(\omega_0^2+  
   \omega_0\omega_{_\text{phys}}^{''}(\eta_m)(\eta -\eta_m)^2 \biggr)\, .
 \eeq
This approximation is well-justified also because, as we said,
backreaction constraints the interval $[\eta_{\ri I}, \eta_{\rii II}]$
to be very small, and therefore $\eta \sim \eta_{m}$ for $\eta, \eta_m
\in [\eta_{\ri I}, \eta_{\rii II}]$, see section \ref{backreactionsec}
for a proof.
Substituting (\ref{omeganearmin}) in (\ref{eqofmoUpot}), the solution is readily found:
 \begin{gather}
  f_\vk(\eta) = B_1 \; W(i \kappa \,\eta_m , \sigma, 2i\kappa\,\eta)
  + B_2 \; W(-i \kappa\, \eta_m , \sigma, -2i\kappa\,\eta)
  \nonumber \\  
  \kappa  \equiv {\sqrt{\omega_0 \omega_{_\text{phys}}^{''}(\eta_m)} \ov H} \qquad  
  \sigma \equiv {\sqrt{9H^2-4\omega_0^2-4 H^2 \kappa^2 \eta_m^2} \ov 2H} \, ,
 \end{gather}
where $W(a, b, z)$ is the Whittacker function.

\subsubsection{Initial conditions, the Bogoljubov coefficients $\alpha_\vk, \beta_\vk, B_1,
 B_2$ and particle creation.}

The result regarding the  coefficients  $\alpha_\vk, \beta_\vk, B_1,
B_2$ in (\ref{piecewisesolution}) is robust, because it is
independent of whatever assumption is made about the behaviour of
$\omega$, in particular, of the
form of $\omega$ assumed in equation (\ref{omeganearmin}). The only
information we need is that the interval $[\eta_{\ri I}, \eta_{\rii II}]$ is
very small because of backreaction (see section
\ref{backreactionsec}) so that
 \beq
  \Delta = {\eta_{\ri I} -\eta_{\rii II} \ov \eta_{\ri I}} \ll 1 \, ,
 \eeq
and that the WKB solutions satisfy the equation (\ref{WKBequation}),
while the partial solutions $\mathcal{U}_{1, 2}$ satisfy
 \beq
  \mathcal{U}_{1, 2}'' + V(\eta, \vk)^2\mathcal{U}_{1, 2} = 0  
  \qquad \eta \in [\eta_{\ri I}, \eta_{\rii II}]\, .
 \eeq
By asking for the continuity of the solution and its first derivative,
we obtain in full generality
 \be \label{Bogolcoeffgener}
  B_1 & = {\mathcal{W}\{\varsigma_\vk \, u_1, \mathcal{U}_2\} \ov \mathcal{W}\{\mathcal{U}_1, \mathcal{U}_2\}}\biggr|_{\eta_{\ri I}} &
  B_2 & = -{\mathcal{W}\{\varsigma_\vk \, u_1, \mathcal{U}_1\} \ov \mathcal{W}\{\mathcal{U}_1, \mathcal{U}_2\}}\biggr|_{\eta_{\ri I}} \\
  \alpha_\vk & = {\mathcal{W}\{B_1 \, \mathcal{U}_1 + B_2 \, \mathcal{U}_2, u_2\} \ov \mathcal{W}\{u_1, u_2\}}\biggr|_{\eta_{\rii II}} &
  \beta_\vk & = -{\mathcal{W}\{B_1 \, \mathcal{U}_1 + B_2 \, \mathcal{U}_2, u_1\} \ov \mathcal{W}\{u_1, u_2\}}\biggr|_{\eta_{\rii II}} \, ,
 \ee
where $\mathcal{W}$ is the Wronskian.
By expanding for small $\Delta$ around $\eta_{\ri I}$, we obtain (again, in full generality)
 \beq \label{alphabetaIII}
  \alpha_\vk = \biggl(1 -i{V(\eta_{\ri I}, k)\,\eta_{\ri I} \ov 2} {\mathcal{Q} \ov V^2}\biggr|_{\eta_{\ri I}} \Delta 
    + O(\Delta^2)\biggr) \, \varsigma_\vk \, , \qquad
  \beta_\vk = \biggl(i{V(\eta_{\ri I}, k)\,\eta_{\ri I} \ov 2} {\mathcal{Q} \ov V^2}\biggr|_{\eta_{\ri I}} \Delta  
    + O(\Delta^2)\biggr)e^{-{2i \ov \epsilon}\Omega_{|_{\eta_{_{\ri I}}}}} \, \varsigma_\vk \, ,
 \eeq
We see that at leading order $\beta_\vk$ is not proportional to small
ratios of scales such as ${H \ov \Lambda}$, but to
the parameter ${\mathcal{Q} \ov V^2}\bigr|_{\eta_{\ri I}}$ signalling the WKB
violation (see equation (\ref{WKBparameteratorderfour})) and
the ratio $\Delta$. The only constraint to the 
value of $\beta_\vk$ is therefore due to the request of
small backreaction.

The Bogoljubov coefficients also depend on the choice of initial
conditions, that is on the pure phase $\varsigma_\vk$. The bispectrum
will be affected by this choice as well, differently from the spectrum,
where only the modulus square of the mode functions enter the
computation and therefore $|\varsigma_\vk| = 1$. In the following, we
will choose $\varsigma_\vk = 1$.

\section{Bispectrum and modifications to the Bunch-Davies
  result}\label{bispectrumsec}  

The three-point function for the comoving curvature perturbation
can be computed in the $\varphi \neq 0$ gauge and then transformed in
the $\zeta \neq 0$ gauge, using equation (\ref{gaugechangezetavarphi}) at
leading order. In the following, when we write a formula in terms of $\zeta \neq 0$
it must therefore be intended that the computation is performed in the 
 $\varphi \neq 0$ gauge and then transformed in the $\zeta \neq
0$ one, where the quantities relevant for observation are more easily read.

In the in-in formalism, one obtains at leading order \cite{Maldacena:2002vr}
 \beq \label{threepoint} 
  \langle \zeta(x_1)\zeta(x_2)\zeta(x_3)\rangle =
  -2 \text{Re}\left( \int^\eta_\ein d \eta' i
  \langle\psi_{\text{in}}|\zeta(x_1)\zeta(x_2)\zeta(x_3)
  H_{(I)}(\eta')|\psi_{\text{in}}\rangle\right)
 \eeq
where $H_{(I)}$ is the interaction Hamiltonian, while $\ein$, 
$|\psi_{\text{in}}\rangle$ are the initial conformal time and state (vacuum). The
standard result in slow-roll inflation is obtained by choosing
$|\psi_{\text{in}}\rangle$ to be
the Bunch-Davies vacuum and $\ein = -\infty$ \cite{Maldacena:2002vr}.

The basic correlator in perturbation theory is given by the Whightman
function
 \beq \label{Wightman}
  \langle \zeta(\eta)\zeta(\eta') \rangle = G_{\vk}(\eta,\eta') \equiv
  {H^2 \ov \phi_{,t}\!^{2}} \frac{f_k(\eta)}{a(\eta)}\frac{f_k^{\ast}(\eta')}{a(\eta')}
 \eeq
We write the bispectrum as \cite{Babich:2004gb}
\beq\label{shape-func}
  \langle
  \zeta_{\vk_1}(\eta)\zeta_{\vk_2}(\eta)\zeta_{\vk_3}(\eta)\rangle
   = \hat F(\vk_1,\vk_2,\vk_3, \eta)
   = (2\pi)^3 \delta(\sum_i \vk_i) \hat B(\vk_1,\vk_2,\vk_3, \eta) \, .
\eeq
 Scale-invariance requires the function $\hat B$ to be
 homogeneous of degree $-6$ and rotational invariance
 makes it a function of only two variables, which can be taken as 
$x_2\equiv {k_2 \ov k_1}$ and $x_3\equiv {k_3 \ov k_1}$. The
 conservation of momentum, with 
 the triangle inequality, forces $x_2+x_3 \geq 1$. Because of the
 symmetry in $x_2$ and $x_3$, it can be further assumed that $x_3\leq x_2$.

Typical configurations in the standard slow-roll/chaotic models are
the local one, where 
 $x_3 \simeq 0$ and $x_2 \simeq 1$, and the equilateral, where
$x_2\simeq x_3\simeq 1$. 
The standard inflationary model with Bunch-Davies vacuum state tends
to produce non-Gaussianities when the modes 
cross the horizon, leading to a three-point function maximized on the
equilateral configuration. If instead the non-Gaussianites receive a
contribution also during the superhorizon evolution, as for example in
the curvaton scenario, they will tend to be maximum on a more local type of 
configurations \cite{Babich:2004gb}. 

In the next two sections we compute the three-point function for two
cases of interactions: a cubic 
coupling without higher power of derivatives acting on the fields, and
a cubic interaction with higher derivatives. Some of the
details of the computation apply to both cases, and we discuss them at
length within the former. We then use that knowledge to analyze the
latter one, which will turn out
to be the most interesting for observations.

\subsection{Cubic scalar interactions}\label{cubicinteractions}

The interacting Hamiltonian at cubic order in the perturbations can be
obtained by expanding the Einstein-Hilbert action on the quasi de
Sitter background \cite{Maldacena:2002vr}.
The contributions to the correlator (\ref{threepoint}) include a connected and some
disconnected parts depending on how we define the curvature perturbation (that
is, if we use non-linear field redefinitions). We will be using a
redefinition leading to the simplest form of the cubic interaction
\cite{Maldacena:2002vr}: 
 \beq \label{HIcubic}
  H_{(I)} = -\int d^{3}x \, a^3 \, ({\phi_{,t} \ov H})^4 \, {H \ov M_{_\text{Planck}}^2}
    \, \zeta'^2 \partial^{-2} \zeta' \, ,
 \eeq
and will be interested in the connected three-point function. In the
following, we indicate with the terms ``three-point function''
and ``bispectrum'' their connected parts\footnote{Therefore, our
  results can be confronted with the observations only after adding
  the disconnected contribution, which in any case do not have
  enhancements and are subleading in slow-roll parameters, see \cite{Maldacena:2002vr}.}. 

Using equation (\ref{HIcubic}), the three-point function (\ref{threepoint})
becomes, in momentum space,
 \begin{multline} \label{bispectrum}
  \!\!\!\!\langle\zeta_{\vk_1}(\eta)\zeta_{\vk_2}(\eta)\zeta_{\vk_3}(\eta)\rangle \!=\! 
   2 \text{Re} \biggl(\!-i (2\pi)^3 \delta(\sum_i \vk_i)\biggl({\phi_{,t} \ov H}\biggr)^4
   \!\!{H \ov M_{_\text{Planck}}^2}\!\! \int^\eta_{\eta_{\text{in}}} d\eta' {{a(\eta')^3} \ov k_3^2} \prod_{i=1}^3
  \partial_{\eta'}G_{\vk_i}(\eta, \eta')
    + \text{permutations}\!\biggr),
 \end{multline}
where $\eta$ is a late time when all modes $k_i$
are outside the horizon after the modified dispersion
relation has become effectively the standard linear one.

Evaluated using equations
(\ref{piecewisesolution}), (\ref{Dcoef}) at the late time $|\eta| \ll
1$, when all three mode functions  
have exited from the horizon in region {\rm IV}, the Green function reads
 \beq \label{derivativeWightmanWKBviol}
  \partial_{\eta'} G_\vk(\eta, \eta') = 
   {H^2 \ov \phi_{,t}^2} {e^{i{\pi \ov 2}} \ov \sqrt{2}}{H \ov k^{3 \ov 2}}(\alpha_\vk-\beta_\vk) 
   \,\biggl({f^*_\vk(\eta') \ov a(\eta')}\biggr)' 
   \simeq {H^2 \ov \phi_{,t}^2} {e^{i{\pi \ov 2}} \ov \sqrt{2}}{H \ov k^{3 \ov 2}}
   \,\biggl({f^*_\vk(\eta') \ov a(\eta')}\biggr)' \, .
 \eeq  

To compute the bispectrum, one needs to divide up the interval of
time integration in equation (\ref{bispectrum}) into the different
regions of validity of the piecewise solution
(\ref{piecewisesolution}) and compute the various terms.
The number of terms is elevated, but many of them are identical up to
permutation of the external momenta and therefore we can reduce them to
some common classes which we now study. By the shortcut expression ``being in
region X'' referred to a Green function depending on $k_i$, we will
mean that for those values of $\eta', k_i$ the function $f_{k_i}(\eta')$ entering
equation (\ref{derivativeWightmanWKBviol}) is the partial
solution of the field equation valid in region X, as shown in equation
(\ref{piecewisesolution}). 

\subsubsection{All Green functions in region {\rm III}}\label{allIII}

We start from the analysis of the contribution to the bispectrum
(\ref{bispectrum}) when
all the three Green functions $G_\vk$ are in
region {\rm III}, which, as we will show, leads to the largest
enhancements. In that region, the function in
(\ref{derivativeWightmanWKBviol}) has  
the form
 \beq
  \partial_{\eta'} G_\vk(\eta, \eta') = \partial_{\eta'} G_\vk^{(\alpha)}(\eta, \eta')+
   \partial_{\eta'} G_\vk^{(\beta)}(\eta, \eta') \, ,
 \eeq
where we have distinguished the positive- and negative-frequency
branches $u_{1, 2}(k, \eta)$ of the 
solution in region {\rm III}, see equation
(\ref{piecewisesolution}). 

There are therefore two possible contributions to the three-point
function: one where the three Whightman function are all
in the same positive or negative branch of the solutions, the other
where one of them is in the opposite branch. It is convenient to
concentrate first on the latter case.

The leading contribution in powers of $\beta_\vk$ to the standard slow-roll result
is then
 \be \label{interferencecorrection}
  \hat F_{3, \beta} & = -2 \text{Re} \biggl[-i (2\pi)^3 \delta(\sum_i \vk_i)
   {H^2 \ov \phi_{,t}^2}
   {H \ov M_{_\text{Planck}}^2} \int^{\eta'_{\riii III}}_{\eta'_{\rii II}}
   d\eta'{1 \ov k_l^2} {e^{i{3 \pi \ov 2}} H^3 \ov \prod_{i=1}^3 (2\,k_i)}
   \beta_{\vk_j}^* \prod_{h \neq j} \gamma^*(\vk_h, \eta') 
     \gamma(\vk_j, \eta')
   \nonumber \\
   & \qquad\qquad\qquad\qquad \times
  \,  e^{i{\Lambda \ov H}\bigl(\Omega(\eta'; k_h)-\Omega(\eta'; k_j)\bigr)}\biggr]+\text{permutations} 
  \qquad h,j \in \{1, 2, 3\} \, .
 \ee
The
Green function depending on $k_j$ is in its negative-frequency branch,
and those depending on $k_h, \, h \neq j$ are in their positive-frequency
one. The limits of integration 
$\eta'_{\rii II}, \eta'_{\riii III}$ are given, respectively, by
the largest and the smallest among
$\eta_{\rii II}^{(k_i)}$ and $\eta_{\riii III}^{(k_i)}$, $i = 1, 2,
3$. We have also defined
 \begin{gather} 
  \label{gammapar}
  \gamma(\vk, \eta) = {1 \ov \sqrt{U}}\biggl({1 \ov 2k}{U' \ov U}+iU+{\mathcal{H} \ov k}\biggr) \, ,
 \end{gather}
and the functions $U(\eta', k) = U(y_k(\eta'))$ and $\Omega(\eta', k) =
\Omega(y_k(\eta'))$ are given in equation
(\ref{Omegaphase}). $\mathcal{H}$ is the Hubble rate in conformal
time, $M_{_\text{Planck}}$ is the reduced Planck mass.

We rewrite (\ref{interferencecorrection}) as
 \beq
  \hat F_{3, \beta}(k_1, k_2, k_3) = A \; \delta \hat{F}_{3, \beta}(k_1, k_2, k_3)
 \eeq
where we factorize the amplitude, including the overall scale dependence,
 \beq \label{standardfactor}
  A = 4 (2\pi)^3 \delta(\sum_i \vk_i){H^6 \ov \phi_{,t}^2 M_{_\text{Planck}}^2}
  {k_1^2k_2^2k_3^2 \ov \prod_{i=1}^3 (2k_i^3)} \sum_l {1 \ov k_t k_l^2} \, ,
  \qquad k_t = \sum_{i=1}^3k_i \, ,
 \eeq
while 
 \be \label{rescaledthreepoint}
  \delta \hat{F}_{3, \beta}(k_1, k_2, k_3) = 
  \sum_j \text{Re} \biggl[{\beta_{\vk_j}^* \ov 2} \int^{\eta_{\riii III}}_{\eta_{\rii II}}
   d\eta' k_t
  \prod_{h\neq j}\gamma^*(\vk_h, \eta')\gamma(\vk_j, \eta')
  \,  e^{i{\Lambda \ov H}\bigl(\Omega(\eta', k_h)-\Omega(\eta', k_j)\bigr)}\biggr] \, ,
 \ee

The amplitude $A$ is precisely the standard
slow-roll result \cite{Maldacena:2002vr}, hence $\delta \hat{F}_{3, \beta}$
is the putative enhancement factor, to which we now turn. Recall
that the modified dispersion relation implies that the frequency
$\omega(\eta, k)$ depends on $k, \eta$ and the scale of new physics
$\Lambda$ as in equation (\ref{powerexp}). We then change the variable $\eta'$
in the integral in equation (\ref{rescaledthreepoint}) to
 \beq \label{changecoordyIII}
   y = {H \ov \Lambda} k_{_\text{max}}\eta' \, , \qquad 
   k_{_\text{max}} \equiv \text{max}(k_1, k_2, k_3)
 \eeq 
so that equation (\ref{rescaledthreepoint})
reads 
 \begin{gather}
  \label{interferenceFourier}
  \delta \hat{F}_{3, \beta}(x_1=1, x_2, x_3) = 
  \sum_j \text{Re} \biggl[{\beta_{\vk_j}^* \ov 2} {\Lambda \ov H}
  x_t \int^{y_{\riii III}}_{y_{\rii II}} dy 
  \prod_{h\neq j}\gamma^*(x_h, y)\gamma(x_j, y)
  \,  e^{i{\Lambda \ov H} S_{\beta}(x_1, x_2, x_3, y)}\biggr],
 \end{gather}
where we have assumed, for instance, $k_1 = k_{_\text{max}}$
and recovered the usual bispectrum variables $x_i = {k_i \ov
  k_{_\text{max}}}$ introduced at the end of 
section \ref{formalismnotation}. To write the formulas in a
more symmetric way, here and in the following, we also indicate formally the ratio 
$x_1 = {k_1 \ov  k_{_\text{max}}}$, although it is equal to
1 because of our assumption. 

We have also defined
 \begin{gather} \label{sumOmegas}
  S_{\beta}(x_1, x_2, x_3, y) \equiv \sum_{h \neq j}\Omega(y, x_h) - \Omega(y, x_j)
  \equiv S_{0\beta}(\{x\}, y)+{H^2 \ov \Lambda^2} S_{2\beta}(\{x\}, y) \\
  S_{0(2)\beta}(\{x\}, y) = \sum_{h \neq j}S_{0(2)}(x_h, y)- S_{0(2)}(x_j, y)  \, .
 \end{gather}
where the dependence of $\Omega$ on $x_i, y$ is due to the fact that 
$y_{k_i}(\eta') \equiv k_i\epsilon \eta' = x_i \, y$, see equation
(\ref{Omegaphase}), and 
$S_{0}, S_{2}$ are the
quantities appearing in the second equation in (\ref{Omegaphase}).

The limits of integration in 
(\ref{interferenceFourier}) are computed as the values for which the
WKB conditions are violated. In general, one finds that 
$y_{\rii II} = {H \ov \Lambda}k_{_\text{max}}\eta^{(k_{\text{max}})}_{\rii II}
\sim -1$, since the corrections to the linear 
  dispersion relation at the time
  $\eta_{\rii II}$ must be quite important in order to drive the
  frequency close to the turning point (see figure \ref{INOUTdispersion}),
  and therefore looking at (\ref{powerexp}) it must be
  ${p_{\text{max}} \ov \Lambda} \sim 1$. The upper limit is instead 
$y_{\riii III} = {H \ov \Lambda}k_{_\text{max}}\eta^{(k_{\text{min}})}_{\riii III}
\sim -{H \ov \Lambda}{1 \ov x_{_{k_{\text{min}}}}}$. 

Evidently, the contribution
(\ref{interferenceFourier}) is sizable if the interval
of integration is sufficiently large. We are therefore interested in
this case, hence we will consider $ x_{_{k_{\text{min}}}} \gtrsim
10 \epsilon$, which 
however for $\epsilon$ reasonably small
leaves plenty of room for all the values of the wavenumber relevant for
the CMBR observations (consider for example that already for the
supersymmetric GUT
scale, it is $\epsilon \lesssim 10^{-3}$ if $H \sim 10^{-5}M_P$). 

The integral in (\ref{interferenceFourier}) is a typical Fourier
integral which can be well approximated by the technique of stationary 
phase, since ${\Lambda \ov H} \gg 1$ \cite{Erdelyi}. As it is known,
its approximated 
solution depends on the critical points of $S_{0 \beta}(\{x\}, y)$ as a
function of $y$. Their presence and nature depend on the
configuration of the external momenta
$k_{1, 2, 3}$, which act as parameters in the function.

We call {\em stationary point of order $n-1$} an interior point $y_*$ such that
 \beq \label{critpointordern}
 \partial_y^n S_{0 \beta}(\{x\}, y)|_{y=y_*} \equiv S_{0 \beta}^{(n)}(\{x\}, y_*) \neq 0, \qquad 
 \partial_y^m S_{0 \beta}(\{x\}, y)|_{y=y_*} = 0\, \qquad \forall\, m < n \, .
 \eeq
The leading order solution to the integral is then \cite{Erdelyi}
\begin{itemize}
 \item[-] if $n \in 2\mathbb{N}$ 
 \begin{multline}  \label{statcriteven} 
  \!\!\!\!\!\!\!\!\!\!\!\!\delta \hat F_{3, \beta} = 
   \text{{\small $\sum_j$}}
   \, \, \biggl({\Lambda \ov H}\biggr)^{1-{1\ov n}}  
   {x_t \, \Gamma\bigl({1 \ov n}\bigr) \ov n \bigl({|S_{0\beta}^{(n)}(\{x\}, y_*)| \ov n!}\bigr)^{{1 \ov n}}}
   \text{Re}\bigl[ \prod_{h\neq j}\gamma^*(x_h, y_*)\gamma(x_j, y_*) \, \beta_\vk^* \, 
   e^{i{\Lambda \ov H} S_{\beta}(\{x\}, y_*)+i{\pi \ov 2n}\text{sign}S^{(n)}_{0, \beta}(\{x\}, y_*)}\bigr]
 \end{multline}
 \item[-] if $n \in 2\mathbb{N}+ 1 > 1$
 \begin{multline}  \label{statcritodd} 
  \delta \hat F_{3, \beta} = 
   \text{{\small $\sum_j$}}
   \, \, \biggl({\Lambda \ov H}\biggr)^{1-{1\ov n}}
   {x_t \, \Gamma\bigl({1 \ov n}\bigr) \ov n \bigl({|S_{0\beta}^{(n)}(\{x\}, y_*)| \ov n!}\bigr)^{{1 \ov n}}}
    \cos\bigl({\pi \ov 2n}\bigr)\text{Re}\bigl[\prod_{h\neq j}\gamma^*(x_h, y_*)\gamma(x_j, y_*) \, \beta_\vk^* \, 
   e^{i{\Lambda \ov H} S_{\beta}(\{x\}, y_*)}\bigr]
 \end{multline}
 \item[-] if $n = 1$~\footnote{\label{Nearcriticalpoints} In fact, if 
$|S_{0 \beta}'(\{x\}, y_{\text{{\riii III}/{\rii II}}})| < 
\sqrt{{H \ov \Lambda}}\bigl({|S_{0 \beta}^{(2)}(\{x\}, y_{\text{{\riii III}/{\rii II}}})| \ov 2}\bigl)^{{1 \ov 2}}$, 
the correct approximation must take into account the
higher order terms in the expansion of $S_{0 \beta}$. A better result
including the second order is then given by (\ref{statcriteven}) 
for $n=2$, evaluated at $y_{\text{{\riii III}/{\rii II}}}$, divided by
-2, and further multiplied by 
$e^{-i{\pi \ov n}\text{sign}S^{(n)}_{0, \beta}(\{x\}, y_{_{\riii III}})}$ 
in the case of $y_{\text{{\riii III}}}$. If necessary,
one can also go to higher orders. In these cases the boundary point
can be called
a nearly critical point.}
 \beq  \label{boundmax}
  \delta \hat F_{3, \beta} = {1 \ov 2}
   \text{{\small $\sum_j$}}
   {x_t \ov S_{0\beta}^{(1)}(\{x\}, y)}
  \text{Im}\bigl[\prod_{h\neq j}\gamma^*(x_h, y)\gamma(x_j, y) \, \beta_{\vk_j}^*  
  e^{i{\Lambda \ov H} S_{\beta}(\{x\}, y)}\bigr]\bigr|^{y_{\riii III}}_{y_{\text{{\rm II}}}}
 \eeq 
\end{itemize}
Observe that $\gamma(x, y)$ does not go to zero if the WKB
approximation is valid.

We recognize the appearance of enhancement factors proportional to
$\bigl({\Lambda \ov H}\bigr)^{1-{1\ov n}}$ in the presence of
stationary points. We postpone the detailed analysis and comments
regarding this to section \ref{generalfeaturesbispectrum}.

\subsubsection{All Green functions in region {\rm III} in their
  positive-energy branches and the case of dispersion relations not
  violating the WKB condition at early times}

The contribution to the bispectrum when all the three Green functions
in formula (\ref{bispectrum}) are in their positive-energy branch can
be obtained by the replacements
 \beq
  \beta_{\vk_j}^* \to \alpha_{\vk}^* \simeq 1 \, ,  \qquad 
  -i\Omega(\eta', k_j) \to i\Omega(\eta', k_j)\, 
  \qquad 
 \eeq 
in equations (\ref{interferencecorrection}) and
successive, which in particular amount to the replacement
 \beq
  S_{\beta}(\{x\}, y) \to S_{\alpha}(\{x\}, y) \equiv 
  S_{0\alpha}(\{x\}, y)+S_{2\alpha}(\{x\}, y) =
   \sum_j S_0(x_j, y) + \sum_j S_2(x_j, y)
 \eeq
in equation (\ref{interferenceFourier}). If we now investigate the presence of
stationary points for the function $S_{0\alpha}(\{x\}, y)$ as we did before
for $S_{0\beta}$, we find that there are none. In fact, let us check the
vanishing of the first derivative, which is a necessary condition to
have a stationary point $y_*$ of any order:
 \beq \label{critpointallposbranchIII}
  \dot S_{0\alpha}(\{x\}, y_*) = \omega(x_1, y_*) 
  + \omega(x_2, y_*) + \omega(x_3, y_*) = 0 \, .
 \eeq
This condition is never satisfied for non-trivial configurations, as
the quantities $\omega(x_{i}, y_*)$ 
are always positive.
The approximated solution for the contribution to 
the three-point function is then
 \beq  \label{allthreeposIII}
  \delta \hat F_{3, \alpha} = {1 \ov 2}\,
   \text{{\small $\sum_j$}} \, {x_t \ov S_{0\alpha}^{(1)}(\{x\}, y)}
  \text{Im}\bigl[\prod_{h\neq j}\gamma^*(x_h, y)\gamma(x_j, y) \,
    e^{i{\Lambda \ov H} S_{\alpha}(\{x\}, y)}\bigr]\bigr|^{y_{\riii III}}_{y_{\text{{\rm II}}}}
 \eeq 

The case of a dispersion relations not violating the WKB condition at
early times follows this same pattern with $\beta_{\vk} = 0$ to start
with, while the time 
integral in the analogous of equation (\ref{interferenceFourier}) is extended
to minus infinity\footnote{The time integration path must be chosen
  such that the oscillating piece in the exponent of the integrand
  becomes exponentially decreasing. This corresponds to taking the
  vacuum of the interacting theory.}. The necessary 
condition for the presence of stationary points is therefore still given by
(\ref{critpointallposbranchIII}), which shows that none
appears and therefore there is no enhancement, in
agreement with what found
in \cite{Ashoorioon:2011eg}. 
 
\subsubsection{Green functions in region {\rm IV}}\label{bispregionIV}

By computing the relevant Green functions, it is seen that
the contribution to the bispectrum when (some of) the Green
functions are in region {\rm IV} can be obtained by
replacing the functions
$\gamma(\vk, \eta)$ and $U(\vk, \eta)$ in equation
(\ref{interferencecorrection}) respectively with $i$ and 
$1$ for the Green function(s) in 
region {\rm IV}. However, since we imagine computing the bispectrum shortly after
the exit from horizon of the modes, because in that case the error involved in
using the relation (\ref{gaugechangezetavarphi}) at leading order is
minimized, this contribution will not be very large.

\subsubsection{One (or more) Green function(s) in region {\rm II}, the others
  in {\rm III}}\label{oneIItwoIII}

We now briefly study the contribution to the bispectrum when one (or more) of the three Green
functions in its formula are in region {\rm II}. For
definiteness, we consider the case where one Green function, say the
one depending 
on $k_3$, is in region {\rm II} and the others are in region  
{\rm III}. It follows that 
 \beq
  k_{1, 2} \leq k_3 \, . 
 \eeq
The other
configurations are readily obtained by permutation.

The generic contribution to the standard bispectrum can then be written as
 \be \label{interferencecorrection32}
  \hat F_{3, B} & = -2 \text{Re} \biggl(i (2\pi)^3 \delta(\sum_i \vk_i)
   {H \ov M_{_\text{Planck}}^2} 
   \int_{\eta^{^{(k_3)}}_{\ri I}}^{\eta^{^{(k_3)}}_{\rii II}} d\eta' {e^{i \pi} H^2 \ov \prod_{i=1}^2 (2\,k_i)}
   (\alpha/\beta)_{\vk_1}^* (\alpha/\beta)_{\vk_2}^*{1 \ov k_l^2}
     \\
   & \qquad\qquad \times \gamma^{(*)}(\vk_1, \eta')
   \gamma^{(*)}(\vk_2, \eta') a(\eta')\partial_{\eta'} G_{k_3}(\eta, \eta')
  \,  e^{\pm {\Lambda \ov H} i\Omega(\eta', k_1)\pm {\Lambda \ov H} i\Omega(\eta', k_2)}\biggr) 
   +\text{permutations} .
   \nonumber
 \ee
where the signs in the phase of the integrand are plus/minus if the
contribution involves $(\alpha/\beta)_{\vk_{1, 2}}^*$, and similarly we have
$\gamma^*/\gamma$ in presence of $(\alpha/\beta)_{\vk_{1, 2}}^*$. 

The three-point function includes now a Green function of the form,
see equation (\ref{derivativeWightmanWKBviol}),
 \beq
  \partial_{\eta'} G_{k_3}(\eta, \eta') =
   {H^2 \ov \phi_{,t}^2} {e^{i{\pi \ov 2}} \ov \sqrt{2}}{H \ov k^{{3\ov 2}}} 
   \,\left({B_1^* \, \mathcal{U}_1^*(\eta', k) + B_2^* \,
     \mathcal{U}_2^*(\eta', k) \ov a(\eta')}\right)' \,.
 \eeq
By using the results in (\ref{Bogolcoeffgener}), we obtain 
 \begin{multline}
   \left({B_1^* \, \mathcal{U}_1^*(k, \eta') + B_2^* \, \mathcal{U}_2^*(k, \eta') \ov a(\eta')}\right)' = 
   {-\mathcal{H}\,u_2(\eta_{\ri I})\;\mathcal{W}\bigl\{\mathcal{U}_1^*(\eta'), \mathcal{U}_2^*(\eta_{\ri I})\bigr\} +u'_2(\eta_{\ri I})\;\mathcal{W}\bigl\{\mathcal{U}_1^*(\eta_{\ri I}), \mathcal{U}_2^*(\eta')\bigr\} \ov a(\eta')\,\mathcal{W}\bigl\{\mathcal{U}_1^*(\eta_{\ri I}), \mathcal{U}_2^*(\eta_{\ri I})\bigr\}} \\
  + {u_2(\eta_{\ri I})\;\mathcal{U}_{[2}^{'*}(\eta_{\ri I}), \mathcal{U}_{1]}^{'*}(\eta') + \mathcal{H}\,u_2(\eta_{\ri I})\;\mathcal{U}_{[2}^{*}(\eta_{\ri I}), \mathcal{U}_{1]}^{*}(\eta') \ov a(\eta')\,\mathcal{W}\bigl\{\mathcal{U}_1^*(\eta_{\ri I}), \mathcal{U}_2^*(\eta_{\ri I})\bigr\}} \, ,
 \end{multline}
where $[ \; ]$ indicates antisymmetrization.

The time integral in equation
(\ref{bispectrum}) is over the interval 
 $ 
 [\eta_{\ri I}^{^{(k_3)}}, \eta_{\rii II}^{^{(k_3)}}] =
  [\eta_{\ri I}^{^{(k_3)}}, \eta_{\ri I}^{^{(k_3)}}(1-\Delta^{^{(k_3)}})] \, ,
 $ 
thus we change variables as 
 \beq \label{varchangesmallDelta} 
  \eta' = \eta_{\ri I}^{^{(k_3)}}(1-\Delta^{^{(k_3)}} v)
 \eeq
and obtain, after expanding for
$\Delta \ll 1$ and recalling that 
$V(\eta_{\ri I}, k_3)^2 \simeq \mathcal{H}_{\ri I}^2$, 
 \beq
   \partial G_{k_3}(\eta\sim 0, v) 
   \simeq e^{i{\pi \ov 4}} {H^2 \ov \phi_{,t}^2} {H \ov k_3^{{3\ov 2}}} 
   {u_2(\eta^{^{(k_3)}}_{\ri I}) \ov \eta^{^{(k_3)}}_{\ri I}} (1+i\Delta^{^{(k_3)}}\, v) \, .
 \eeq

We now write 
 \beq \label{factorstandardbispectrumBpart}
  \hat F_{3, B} = A \, \delta\hat F_{3, B} + \text{permutations} \, , 
 \eeq
where $A$ is defined in (\ref{standardfactor}).
We also change the variable 
$y_{k_i}'$ in (\ref{Omegaphase}) to 
$\xi = {y_{k_i}(\eta_{\ri I})-y_{k_i}' \ov
  \Delta y_{k_i}(\eta_{\ri I})}$ so that the
equation is more naturally written in terms of the
variable $v$ defined in (\ref{varchangesmallDelta}) as
 \beq   
  {\Lambda \ov H}\Omega(y_{k_i})  \; \to  \;
   -\eta_{\ri I}\Delta \int^{{\eta_{\ri I}-\eta' \ov \eta_{\ri I}\Delta}} k_i U(\xi, k_i) \,  d\xi 
  =
  -\eta_{\ri I}\Delta \int^{v}k_i U(xi, k_i) \,  d\xi 
   \equiv - \eta_{\ri I} \, \Delta \, \Omega(v, k_i)\, ,
 \eeq 
and equations (\ref{interferencecorrection32}),
(\ref{factorstandardbispectrumBpart}) give
 \begin{multline} \label{interferenceFourierII}
  \delta\hat F_{3, B} = \text{Re}\biggl[{\Delta \ov 2}^{^{(k_3)}}
   e^{-i {\pi \ov 4}-i\eta^{^{^{\!(k_3)}}}_{\ri I}\Delta^{^{^{\!(k_3)}}}\Omega(k_3, 0)}
   \int^{1}_{0} dv
   (\alpha/\beta)_{\vk_1}^* (\alpha/\beta)_{\vk_2}^*    
   \gamma^{(*)}(\vk_1, v) \gamma^{(*)}(\vk_2, v)
   \\
   \times  (1+{k_2 \ov k_3}+{k_1 \ov k_3})(1+i\Delta^{^{(k_3)}} v) 
   e^{i\eta_{\ri I}\Delta\Sigma_{B}(k_1, k_2, v)}\biggr] \, ,
 \end{multline}
where
 \beq
  \Sigma_{B}(k_1, k_2, v) = \mp \Omega(v, k_1) \mp  \Omega(v, k_2) \, ,
  \qquad \mp \leftrightarrow (\alpha/\beta)_{\vk_{1, 2}}^* \, .
 \eeq
We see that now the asymptotic stationary phase
approximation of the Fourier integral is not accurate, as there is no
large factor in the phase of the integrand (since 
$\Delta^{^{(k_3)}} \ll 1$). Therefore, also when the combination of 
signs leads to interference and phase cancellation, there will appear
no enhancement factor. Observe also that the correlator is proportional
to the small factor $\Delta^{^{(k_3)}} \ll 1$. 

It follows that, overall, 
this contribution is suppressed compared to 
the one we discussed in section \ref{allIII}. It is straightforward to
realize that when more of the Green functions entering equation
(\ref{bispectrum}) are in region {\rm II}, the
contribution is also suppressed. 
This result is very robust, because it is not based on the specific form of
the solutions $\mathcal{U}_1(\eta, k)$: we have in fact only used
that $\Delta^{^{(k_3)}} \ll 1$ and the general form
(\ref{Bogolcoeffgener}) of the 
coefficients $B_{1, 2}$ which is valid for all solutions.

\subsubsection{One (or more) Green functions in region {\rm I}}\label{oneItwoIII}

The last kind of contributions that we are left to analyze is the one
where one or more of the Green functions in the formula for the
bispectrum are in region {\rm I}. There can be various possibilities,
depending on what region the other Green functions belong to at those
times, given $V(\eta, k)$.
If (some of) the other Green functions are in region {\rm II}, we expect the
contribution to be suppressed in a way similar to that
discussed in section \ref{oneIItwoIII}. Recall that in region {\rm I}, the WKB solution
has only the positive-energy branch $u_1$ because we chose the
adiabatic vacuum, see
(\ref{piecewisesolution}), therefore no interference is possible between
functions in region {\rm I} only. If instead at least one of the Green
functions is in 
region {\rm III}, we could conceive the presence of
interference terms possibly leading to enhancement factors\footnote{For this,
  at least one of the Green functions in region {\rm III} must be in the negative
  frequency branch of the solution,  
  otherwise there will be no enhancement, as discussed in section
  \ref{allIII}.}.  

Let us consider this case in more details, imagining for definiteness
that the Green function in region {\rm I} is the one depending on
$k_3$, while the others are in region {\rm III}. The form of the 
contribution to the bispectrum is the same as that in equation 
(\ref{interferencecorrection}), with the difference that now
$\omega(\eta, k_3), \Omega(\eta, k_3)$ and $\gamma(k_3, \eta)$ feel, through equation
(\ref{powerexp}),
effects that are not suppressed in region {\rm I}. 

Following the derivation in section \ref{allIII}, in order to have the 
enhancements there must exist momenta configurations allowing the
existence of stationary points $y_*$ for the function 
given by the sum of phases of 
the WKB solutions multiplied by the large factor ${\Lambda \ov H}$. 
In the situation we are discussing here, it however appears more difficult to
meet this requirement.
Indeed, let us consider just the first and necessary condition
for having stationary points of order at least $1$: the equation is, say,
 \beq \label{minenhanctwoIIIoneI}
  \omega(x_1, y_*) - \omega(x_2, y_*)+
  \omega(x_3, y_*) = 0 \, .
 \eeq
We expect that satisfying this equation for
non-trivial configurations 
would be more difficult  than in the case of section \ref{allIII},
because the frequencies are now in different parts of their curves:
$\omega(x_{1,2})$ in region {\rm III}, 
while $\omega(x_{3})$ in region
{\rm I}. Therefore, we expect not to find enhancements.

However, since this obstacle to having large enhancements strictly
depends on the form of the frequencies $\omega$, there
could be cases where the conditions for the stationary points can be
satisfied. As this is strongly 
model-dependent, we will not deal with this point
further, and leave it to be addressed case by case in the models of
interest. 

\subsubsection{Comments on the features of the corrections to the slow-roll bispectrum
  from Lorentz and WKB violation at early
  times}\label{generalfeaturesbispectrum} 

It is convenient to pause for a moment and comment in general terms
the results found so far.
The most notable features of the bispectrum from cubic interactions in
the case of Lorentz and 
WKB violation at early times 
emerge from the equations (\ref{statcriteven}), (\ref{statcritodd}),
(\ref{boundmax}) and (\ref{allthreeposIII}), as we have shown in sections
\ref{bispregionIV}, \ref{oneIItwoIII}, \ref{oneItwoIII} that the 
other contributions are suppressed.

The contribution given by equation (\ref{allthreeposIII}), multiplied
by the factor in (\ref{standardfactor}), matches
the standard result, from which it deviates only
for the presence of tiny superimposed oscillations. The same
qualitative behaviour of the bispectrum occurs for dispersion
relations not violating the WKB condition at early times, which
therefore do not show strong deviations from the standard slow-roll
results. 

The corrections to the standard result for dispersion relations that
do violate the WKB condition at early times are instead quite different.
They present a modulation in wavenumber space, descending
from the oscillatory behaviour of (\ref{statcriteven}), (\ref{statcritodd}),
(\ref{boundmax}), but in addition the contribution of certain
configurations can also 
be enhanced, depending on the form of the dispersion relation.
These configurations are
individuated by the system (\ref{critpointordern}), and can therefore
be found by the knowledge of the dispersion relation\footnote{Also
  near-to-critical points lead to enhancements, which are however
  smaller than those of critical points, see footnote
  \ref{Nearcriticalpoints}.}. Compared to the Bunch-Davies slow-roll
result, they lead to the enhancement factors 
 \beq \label{enhancementfactors} 
 {\hat F_{3, \beta} \ov \hat F_{_\text{BD}}} = \delta \hat F_{\beta} 
  \sim \sum_j \biggl({\Lambda \ov H}\biggr)^{1-{1 \ov n}}
   {1 \ov |S_{0\beta}^{(n)}(\{x\}, y_*)|^{{1\ov n}}}|\beta_{\vk_j}^*| \, ,
 \eeq
which can be large for $n > 1$.

Let us compare this result with the one found in
\cite{Holman:2007na} and anticipated in \cite{Chen:2006nt}, based on
the modified vacuum approach, which models the transplanckian
  physics imposing a cutoff on the theory in momentum space. 
The result in that case also showed
oscillations in the bispectrum and enhancements of the order
of ${\Lambda \ov H}$, but only for the enfolded configuration. 
We see that in general the enhancements in the case of modified
dispersion relation have a smaller magnitude: it would be 
the same if there were configurations
$\{x\}_{\text{enh}}$, such that there
existed a stationary point $y_*$ with the property that all 
derivatives of $S_{0 \beta}(\{x\}_{\text{enh}}, y)$ at $y_*$ were
zero. The enhancement factor 
would then be given by the limiting behaviour of
(\ref{enhancementfactors}) for $n \to \infty$.
In fact, this occurs for example if
 \beq \label{SphaseHT}
  S_{0 \beta}(\{x\}, y)=S_{HT}(\{x\}, y)=(\sum_{h \neq j}x_h-x_j)(y-y_c), \qquad
  y_c= {H \ov \Lambda} \text{max}(k_1, k_2, k_3)\eta_c \, , 
 \eeq
which is indeed the phase of the integrand in the bispectrum
formula obtained via the modified vacuum approach,
where $\eta_c$ 
is the time cutoff corresponding to the appearance of 
new physics \cite{Holman:2007na}. In this case, all
derivatives of $S_{HT}(\{k\}, x)$ are zero except the first, which is
zero for the enfolded configurations 
given by $\sum_{h \neq j}k_h-k_j = 0$~\footnote{In this case we cannot
talk of stationary points, since they are not isolated: the derivative
is zero over the whole range of $y$.}. However, from the point of view of
modified dispersion relations, where $\omega \neq k$, it appears that such an
eventuality would be rather peculiar.

The final magnitude of the enhancement depends on the value of the
Bogoljubov parameter $\beta_\vk$. In section \ref{backreactionsec}
we will show how the constraints from backreaction affect it and
reduce the magnitude of the overall enhancement factor
(\ref{enhancementfactors}). However, by looking at equation
(\ref{interferenceFourier}), we spot that larger enhancements would
appear for interactions that scale with higher powers of 
${1 \ov a(\eta)}$. We turn now to one such example.
 
\subsection{Higher derivative interactions}\label{highderinteractions}

An example of interaction that scales with larger powers of ${1 \ov
  a(\eta)}$ than the cubic coupling (\ref{HIcubic}) is
\cite{Holman:2007na, Creminelli:2003iq}
 \beq 
  \mathcal {L}_{(I)} =\sqrt{-\text{det}g} 
   {c \ov 8 M^4}((\nabla \phi)^2)^2 \, .
 \eeq

Once expanded in perturbations, this term leads to
one contribution quadratic in the fields and to another which is
cubic. The first one has the effect of modifying the sound speed of
the modes in the field equation (\ref{eqofmoUpot}). We neglect this
and focus instead
on the three-point function generated by the cubic
contribution to the Lagrangian and, consequently, to the interacting
Hamiltonian. It can be found that the latter is \cite{Holman:2007na}
 \beq \label{hamilthigder}
  {H}_{(I)} = -\int d^3x a {c \, \phi_{,t} \ov 2 \Lambda^4} 
   \varphi' 
   \bigl({\varphi'}^2-\bigl(\partial_i \varphi \bigr)^2\bigr) 
    = -\int d^3x a {c \, \phi_{,t}^4 \ov 2 H^3 \Lambda^4} \zeta' 
    \bigl({\zeta'}^2-\bigl(\partial_i \zeta \bigr)^2\bigr),
 \eeq
From equation (\ref{threepoint}), we obtain at leading order the correlator
 \be \label{threepointhighder}
  \hat F(k_1, k_2, k_3) & = 2\text{Re}\biggl[ i (2\pi)^3
  \delta^{(3)} \bigl(\sum_i \vec{k}_i\bigr) {c \phi_{,t}^4 \ov 2H^3 \Lambda^4}
  \int_{\eta_{\rii II}}^{\eta_{\riii III}} d\eta' \, a
  \bigl(\partial_{\eta} G_{k_1}(\eta, \eta') \partial_{\eta} G_{k_2}(\eta, \eta')\partial_{\eta} G_{k_3}(\eta, \eta') + \nonumber \\
  & \qquad + \vk_1\cdot \vk_2 G_{k_1}(\eta, \eta') G_{k_2}(\eta, \eta')\partial_{\eta} G_{k_3}(\eta, \eta'))+\text{permutations}
   \biggr]  
    \equiv A \; \delta \hat F_{\partial}\, ,
 \ee
where $A$ is defined in equation (\ref{standardfactor}).

From the
analysis in section \ref{cubicinteractions}, we have learnt 
that the largest correction to the bispectrum occurs when all
the Green functions are in region {\rm III} and one among them is in
the opposite frequency branch with respect to the others. We 
therefore focus on that case.
At leading order in $\beta_\vk$, 
 \begin{gather} \label{enhanchigder} 
  \delta \hat F_{\partial,\, \beta} 
  =  -\text{{\small $\sum_j$}} \text{Re}\biggl[ \, 
  C \, \biggl({\Lambda \ov H}\biggr)^3 
  \int_{y_{\rii II}}^{y_{\riii III}} dy y^{2} \, \beta_{\vk_j}
  e^{i{\Lambda \ov H} S_{\beta}(\{x_h\}, x_j, y)} \, 
  P(\{x_h\}, x_j, y) \qquad h \neq j
    \nonumber \\
  P(\{x_h\}, x_j, y) =\biggl( 6  \prod_{h \neq j} \gamma^*(x_h, y)\gamma(x_j, y)
  + 2 {\vec x_{j+1} \cdot \vec x_{j+2} \, \gamma(x_j, y) \ov x_{j+1}\,x_{j+2}\,\sqrt{U(x_{j+1})U(x_{j+2})}}
  + 2 {\vec x_{j+1} \cdot \vec x_j \, \gamma^*(x_{j+2}, y) \ov x_{j+1}\,x_j\,\sqrt{U(x_{j+1})U(x_j)}}
    \nonumber \\
  \qquad \qquad +  2 {\vec x_{j+2} \cdot \vec x_j \,\gamma^*(x_{j+1}, y) \ov x_{j+2}\,x_j\,\sqrt{U(x_{j+2})U(x_j)}} 
  \biggr)\biggr] \, ,
 \end{gather}
where we have changed variables as in equation
(\ref{changecoordyIII}), written 
$\vec x_i \equiv {\vk_i \ov k_{_\text{max}}}$ and used $\gamma, S_\beta$ defined in 
equations (\ref{gammapar}), (\ref{sumOmegas}), while
$C \equiv   {x_t \, \prod_{l=1}^3 x_l^2 \ov \sum_{l > i} x_i^2x_l^2}
  {c H^2 M_{_\text{Planck}}^2 \ov 4 \Lambda^4}$. We have not
indicated the $y$-dependence of $U(x_i, y)$ to avoid cluttering the formula.
In the expression for $P(\{x_h\}, x_j, y)$, indices are defined modulo 3.

Now, we need to study the presence of stationary points for the function
$S_{0 \beta}$, as done in
section \ref{allIII}. We need also to
check the behaviour of 
$P(\{x_h\}, x_j, y)$. In fact, it could go
to zero or be very suppressed for the configurations for which there
exist stationary points of $S_{0 \beta}$. This is what happens in the modified
vacuum case, where the highest power of ${1 \ov a}$ in this
bispectrum contribution is exactly zero on the enhanced folded
configurations, and therefore the final enhancements is reduced
\cite{Holman:2007na}.   
 
It is straightforward to observe that in the case of modified
dispersion relations, $P(\{x_h\}, x_j, y)$ does not generally go to zero nor
it is strongly suppressed. Let us consider the case 
of a stationary/boundary point at the value $y = y_* \neq 0$. Since 
$|y_*| \leq |y_{\rii II}| \sim 1$ we can expand\footnote{The expansion
  is not well 
justified if $y_* = y_{\rii II}$, but even in that case one can check
that there is no cancellation of
$P(\{x_h\}, x_j, y_*)$ in general. In the following, recall also
that ${\epsilon \ov x_i y_*} < 1$ for all $y_* \in [y_{\rii II},
  y_{\riii III}]$.} $P$ in powers of $y_*$: 
 \beq
  \gamma(x_i, y_*) \sim i + i c_i x_i y_* +
   \epsilon \bigl({c_i \ov 2}-{1 \ov x_i y_*}\bigr) + \ldots
 \eeq
where $c_i = \partial_y U(x_i, y)|_{y=y_*} \sim O(1)$.
It follows that\footnote{In equation (\ref{Pfactexpan}), indices with
  the letter $h$ are defined modulo 2, indices with the letters $j$
  or $i$ or $m$ are defined modulo 3.}
 \beq \label{Pfactexpan} 
  i P(\{x_h\}, x_j, y) \sim \kappa_{_{0,0}}+ \kappa_{_{1,-1}}{\epsilon \ov y_*} +
  \kappa_{_{0,1}} \, y_* + \kappa_{_{10}} \, \epsilon + \kappa_{_{11}} \, \epsilon \, y_*
  + \ldots
 \eeq
{\small \begin{gather}
  \kappa_{_{0,0}} = {(\text{{\small $\sum_{h \neq j}$}} x_h-x_j)(x_t^2 - 4 x_{j+1} x_{j+2}) \ov x_1 x_2 x_3} 
   \qquad 
  \kappa_{_{1,-1}} = -6 i (\text{{\small $\sum_{h \neq j}$}} x_h^{-1}-x_j^{-1})+
    i \text{{\small $\sum$}}_{i=1}^3 x_ix^{-1}_{i+1}x^{-1}_{i+2}
   \nonumber \\
  \kappa_{_{0,1}} = 6 \text{{\small $\sum\limits_{i=1}^3$}} c_ix_{i} 
    -(\text{{\small $\sum_{h \neq j}$}} c_hx_hx^{~}_{\{h+1}x_{h+2\}}^{-1}-c_jx_jx^{~}_{\{j+1}x_{j+2\}}^{-1}) 
    +(\text{{\small $\sum_{h \neq j}$}} c_hx_h^3x_{h+1}^{-1}x_{h+2}^{-1}-c_jx_jx_{j+1}^{-1}x_{j+2}^{-1}) 
     \qquad \nonumber \\
    +\text{{\small $\sum_{h \neq j}$}} x_h(-c_{h+1}+c_j)+x_j\text{{\small $\sum_{h \neq j}$}} c_h
    +\text{{\small $\sum_{h \neq j}$}}
    {c_hx_h \ov 2x_1x_2x_3}(x^2_h(x_{h+1}-x_j)-x_{h+1}^3+x_j^3)+{c_jx_j \ov 2x_1x_2x_3}(x_j^2\text{{\small $\sum_{h \neq j}$}}x_{h}- 
      \text{{\small $\sum_{h \neq j}$}}x_{h}^{3})
   \nonumber \\
  \kappa_{_{1,0}} = 3i(\text{{\small $\sum\limits_{h \neq j}$}} c_i-c_j) 
   -6i(\text{{\small $\sum_{h \neq j}$}}c_hx_h(x_{h+1}^{-1}-x_{j}^{-1}) + c_jx_j(x_{j+1}^{-1}+x_{j+2}^{-1}))
   -\text{{\small ${1 \ov 2}$}}\text{{\small $\sum\limits_{i=1}^3$}} (c_ix^{~}_{\{h+1}x_{h+2\}}^{-1}
     +c_ix_i^2x_{h+1}^{-1}x_{h+2}^{-1})
    \nonumber
 \end{gather}
 \begin{multline}
   \kappa_{_{1,1}} = 3i(\text{{\small $\sum\limits_{h \neq j}$}} c_hx_h(x_{h+1}-x_j)+c_jx_j(x_{j+1}+x_{j+2}))
    +6i (c_{j+1}c_{j+2}x_{j+1}x_{j+2}x_j^{-1}-c_{j}c_{j+1}x_{j}x_{j+1}x_{j+2}^{-1} \qquad
    \nonumber \\
    -c_{j}c_{j+2}x_{j}x_{j+2}x_{j+1}^{-1})
    +\text{{\small ${i \ov 4}$}}
    (\text{{\small $\sum\limits_{i=1}^3$}}x_ic_{i+1}c_{i+2}
    +\text{{\small $\sum\limits_{i > m}$}}{c_ic_mx^3_{\{i}x^{~}_{m\}} \ov x_1x_2x_3})
     \, ,\nonumber 
 \end{multline}}
where $\{ \; \}$ indicates symmetrization.

The coefficient $\kappa_{_{0,0}}$ matches the result of
\cite{Holman:2007na} as expected, and it is indeed zero for the
enfolded configuration. In general, however, in the case of modified
dispersion relation the enhanced configuration need not be the
folded one, and therefore $\kappa_{_{0, 0}} \neq 0$. Even in the case of
the enfolded configuration, the
other coefficients of the expansion will not be zero in general and
need not be utterly small.

Therefore, for a stationary point $y_*$ of
order $n-1$, we obtain \cite{Erdelyi}
\begin{itemize}
 \item[-] if $n \in 2\mathbb{N}$ 
 \begin{multline}  \label{statcritevenhigder} 
  \!\!\!\!\!\!\delta \hat F_{\partial,\, \beta} = 
  -\text{{\small $\sum_j$}} C
   \,\biggl({\Lambda \ov H}\biggr)^{3-{1\ov n}}   
   {y_*^2 \, \Gamma\bigl({1 \ov n}\bigr) \ov n \bigl({|S_{0\beta}^{(n)}(\{x\}, y_*)| \ov n!}\bigr)^{{1 \ov n}}}
   \text{Re}\bigl[P(\{x_{h \neq j}\}, x_j, y_*)\, \beta_\vk^* \, 
   e^{i{\Lambda \ov H} S_{\beta}(\{x\}, y_*)+i{\pi \ov 2n}\text{sign}S^{(n)}_{0, \beta}(\{x\}, y_*)}\bigr]
 \end{multline}
 \item[-] if $n \in 2\mathbb{N}+ 1 > 1$
 \begin{multline}  \label{statcritoddhigder} 
  \delta \hat F_{\partial,\, \beta} = 
  -\text{{\small $\sum_j$}} C
   \,\biggl({\Lambda \ov H}\biggr)^{3-{1\ov n}}
   {y_*^2 \, \Gamma\bigl({1 \ov n}\bigr) \ov n \bigl({|S_{0\beta}^{(n)}(\{x\}, y_*)| \ov n!}\bigr)^{{1 \ov n}}}
    \cos\bigl({\pi \ov 2n}\bigr)
   \text{Re}\bigl[P(\{x_{h \neq j}\}, x_j, y_*)\, \beta_\vk^* \, 
   e^{i{\Lambda \ov H} S_{\beta}(\{x\}, y_*)}\bigr]  \, .
 \end{multline}
\end{itemize}

We conclude that, in the case of modified dispersion relation, the
enhancement for higher derivative interactions, when
present, is actually {\em larger} than 
the one obtained via the modified vacuum approach to transplanckian
physics, if
 \beq
  \biggl({\Lambda \ov H}\biggr)^{1-{1\ov n}} y_*^2 > 1 \, ,
 \eeq
as can be seen comparing with equations (3.31), (3.32) in
\cite{Holman:2007na}. We comment on the likelihood of this at the end
of the paper.
Finally,
\begin{itemize}
 \item[-] if there is no stationary point, $n =1$~\footnote{If 
$|S_{0 \beta}'(\{x\}, y_{\text{{\riii III}/{\rii II}}})| < 
\sqrt{{H \ov \Lambda}}\bigl({|S_{0 \beta}^{(2)}(\{x\}, y_{\text{{\riii III}/{\rii II}}})| \ov 2}\bigr)^{{1 \ov 2}}$, 
 the correct approximation must take into account the
higher order terms in the expansion of $S_{0 \beta}$. In this case, a
better result including the second order is 
then given by (\ref{statcritevenhigder}) 
for the case $n=2$, 
evaluated at $y_{\text{{\riii III}/{\rii II}}}$, and then divided by
-2 and further
multiplied by $e^{-i{\pi \ov n}\text{sign}S^{(n)}_{0, \beta}(\{x\}, y_{\riii III})}$ 
in the case of $y_{\text{{\riii III}}}$. If necessary,
one can also go to higher orders. However, observe that
$y_{\text{{\riii III}}}$ is such that certainly 
$\bigl({\Lambda \ov H}\bigr)^{2} y_{\text{{\riii III}}}^2 \sim 1$,
and therefore there is truly no enhancement at that boundary point.}
 \beq  \label{boundmaxhigder}
  \delta \hat F_{\partial,\, \beta} = - {1 \ov 2}
  \text{{\small $\sum_j$}} C
   \,\biggl({\Lambda \ov H}\biggr)^{2}   
   {y^2 \ov S_{0\beta}^{(1)}(\{x\}, y)}
  \text{Im}\bigl[P(\{x_{h \neq j}\}, x_j, y)
   \, \beta_{\vk_j}^* \, 
  e^{i{\Lambda \ov H} S_{\beta}(\{x\}, y)}\bigr]\bigr|^{y_{\riii III}}_{y_{\text{{\rm II}}}}
 \eeq 
\end{itemize}
Schematically, the enhancement is given by
 \beq \label{enhancementfactorshigder}
  {\hat F_{\partial,\, \beta} \ov \hat F_{\text{BD}}}
   \simeq \sum_j \biggl({\Lambda \ov H}\biggr)^{3-{1 \ov n}}
   {y_*^2 \ov |S_{0\beta}^{(n)}(\{k\}, y_*)|^{{1\ov n}}}|\beta_{\vk_j}^*| \, .
 \eeq

\section{Backreaction}\label{backreactionsec} 

The energy density produced in connection with particle creation 
backreacts on the background. When interactions are turned on, this
energy can be divided in a ``free'' and an ``interaction'' parts.

The study of the backreaction in the case of modified dispersion relations
has so far only dealt with the free part, as in
\cite{Lemoine:2001ar, Brandenberger:2004kx}. The interaction one
has instead been studied mostly
within the approach of modified initial vacuum in
\cite{Greene:2004np} and especially in 
\cite{Holman:2007na}. We will show that the general results found in
\cite{Holman:2007na} are similar to those we find for the case of
modified dispersion relations.

We start with a review of the features of the free part of the
produced energy 
density. We consider the energy generated by the particle creation in
region {\rm III}, which is the most relevant for our results. The
formula for the average energy density can be 
obtained computing the energy-momentum tensor of the inflaton from the typical
Lorentz-breaking action we report in appendix \ref{lorbreackscalact} \cite{Lemoine:2001ar}:
 \beq  \label{freeeneffdisprel}
  \langle 0| \hat\rho | 0 \rangle = {1 \ov 4\pi^2a^4}\int dk\, k^2
   \left[a^2\left|\left({f_k \ov a}\right)'\right|^2
  +\omega^2(k)\left|f_k\right|^2\right],
 \eeq
Using the third line of the solution (\ref{piecewisesolution}), valid
in region {\rm III}, one obtains
 \be
  \langle 0 | \hat\rho | 0 \rangle (\eta ) & =
  {1 \ov 4\pi^2a^4}\int dk k^2\biggl\{
  {1 \ov 2 k \, U} \biggl[\omega ^2 +|g|^2\biggr]+
  {|\beta_k|^2 \ov k\, U}
  \biggl[\omega ^2 + |g|^2\biggr]
  +{\alpha_k\beta_k^* \ov 2k\,  U}\biggl[\omega^2 +g^2\biggr]
   e^{-2i\Omega(\eta, k)} \nonumber \\
   & \qquad\quad
   + {\alpha_k^*\beta_k \ov 2k\,  U}\biggl[\omega ^2 +(g^*)^2\biggr]
   e^{2i\Omega(\eta, k)}\biggr\},
 \ee
where the momentum integral extends over values
for which the solution of region {\rm III} is valid.
Here, $g \equiv {1 \ov 2}{U' \ov U}+ik\, U+\mathcal{H} \sim
i\omega(k, \eta)$, being in region {\rm III}, so that for modified
dispersion relations the terms with
exponentials cancel at leading order even without doing the
integration (averaging) over the oscillating 
exponential, which, as argued in \cite{Holman:2007na}, would damp those
contributions anyway.

The formula for the ``free'' energy density is therefore reduced to
 \beq \label{rhoreduc}
  \langle 0 |\hat\rho| 0 \rangle \simeq {1 \ov 4\pi^2a^4}\int dk\, k^2
  \left({1 \ov 2} + |\beta_k|^2 \right)\omega(k)
  = {1 \ov 4\pi^2}\int dp\, p^2
  \left({1 \ov 2} + |\beta_k|^2 \right)\omega_{_\text{phys}}(p) \, ,
 \eeq
in terms of the physical momentum 
 $ 
   p = {k \ov a} 
 $
 and frequency   
 $   
   \omega_{_\text{phys}}(p) \!=\! {\omega(k) \ov a} = p \text{{\small $F\bigl(-{p \ov \Lambda}\bigr)$}} 
   \, ,
 $ 
see equation (\ref{powerexp}).

We now substitute the expression (\ref{alphabetaIII}) for the
Bogoljubov parameter and change variables as 
${p(\eta) \ov \Lambda} \equiv y(\eta) $, so that the integral
(\ref{rhoreduc}), after discarding the zero-point 
contribution as usual\footnote{In a gravity theory this is a
  non-trivial step, but we uniform ourselves to what usually done in the
  literature on the subject \cite{Lemoine:2001ar,
    Brandenberger:2004kx}.}, yields 
 \beq
  \langle 0 |\hat\rho| 0 \rangle \sim \Delta^2
  \Lambda^4 \, .
 \eeq
This has to be compared to the background energy density $\rho_b = 3\,
H^2 \, M_{_\text{Planck}}^2$, yielding the constraint
 \beq \label{energynobackreaction}
  \Delta < {H \, M_{_\text{Planck}} \ov \Lambda^2}
 \eeq
to avoid issues with the backreaction. It ought to be required that the
minimum $\omega_0$ of $\omega_{_\text{phys}}$ at early times (see figure
\ref{INOUTdispersion}) be different from zero, otherwise,
as discussed in \cite{Brandenberger:2004kx, Danielsson:2004xw}, there
would be WKB violation even at present times and the constraint would
involve $H_{\text{today}}$, becoming very constrictive. If instead
$\omega_0 > 0$, it can be shown that there is 
no further backreaction issue past a certain time
after the end of inflation, because the Hubble rate
decreases and the WKB violation does
not occur any more after the time when $H_{\omega_0} \sim
\omega_0$. In our estimates in this section, we make the assumption
that $H_{\omega_0} \sim H$ at inflation for simplicity, but the general case can
certainly be considered. 

We also have to preserve the slow-roll conditions. We define the
standard slow-roll parameters and link them to the energy and pressure
density via the equations
 \beq
  {d H \ov dt} =- \varepsilon H^2=-{1 \ov 2M_{_\text{Planck}}^2}(\hat p+\hat \rho) \, ,
  \qquad {d^2 H \ov dt^2} =2\varepsilon \mu H^3 
   =-{1 \ov 2M_{_\text{Planck}}^2}\biggl({d \hat p \ov dt}-3H(\hat p+\hat \rho)\biggr), 
 \eeq
where we have called $\mu = \eta_s-\varepsilon$, and $\eta_s$ is the slow-roll
$\eta$-parameter.

Using the formula \cite{Lemoine:2001ar}
 \beq \label{presseffdisprel}
  \langle 0 |\hat p | 0 \rangle = {1 \ov 4\pi^2a^4}
   \int dk \, k^2
  \left[a^2\left|\left({f_k \ov a}\right)'\right|^2 +
  \left({2 \ov 3}k^2 {d\omega^2 \ov dk^2}-\omega^2\right)\left|f_k\right|^2\right].
 \eeq
we obtain  $\langle \hat p \rangle \sim  \Delta^2
  \Lambda^4$ and $\langle {d \hat p \ov dt} \rangle \sim  \Delta^2 H
  \Lambda^4$, from which the bounds 
 \beq \label{slowrollnobackreaction}
  \Delta < \sqrt{\varepsilon} {H M_{_\text{Planck}} \ov \Lambda^2}
  \, , \qquad
  \Delta < \sqrt{\varepsilon |\mu|} {H M_{_\text{Planck}} \ov \Lambda^2}.
 \eeq

We now discuss the interaction contribution to the energy
density. The expectation value of the 
energy-momentum tensor at leading order is given by
 \beq
  \langle T_{\mu\nu}(\vec x, \eta) \rangle = -2\text{Re}\left(i
  \int_{\eta_{\text{{\rm II}}}}^{\eta_{\riii III}}\langle T^{(I)}_{\mu\nu}(\vec x, \eta) H_{(I)}\rangle\right)
 \eeq
where $H_{(I)}$ is given in (\ref{HIcubic}). The energy-momentum tensor
$T^{(I)}_{\mu\nu}$ is also in the interaction picture, and
therefore, one finds that the energy density component is given by the
Hamiltonian density in the interaction picture. Thus, we obtain
 \begin{multline} 
  \langle \hat \rho^{(3)} \rangle  \approx 
  \text{Re}
  \biggl[{3 \ov 2} (2\pi)^3\delta(\sum_i \vk_i)
  \biggl({\phi_{,t} \ov H}\biggr)^2 {H^2 \ov M_{_\text{Planck}}^4 a(\eta)^4}
  \int {d^3k_1 \ov (2\pi)^{{3 \ov 2}}}\int {d^3k_2 \ov (2\pi)^{{3 \ov 2}}}
  \int_{\eta'_{\text{{\rm II}}}}^{\eta'_{\riii III}}d\eta'
    \text{{\small $\sum\limits_{h,j =1}^3$}}{k_1\,k_2\,k_3 \ov k_{h}^2 k_{j}^2}\prod_{i=1}^3 \\
   \qquad \biggl(|\alpha_{\vk_i}|^2 \gamma(k_i, \eta)\gamma^*(k_i, \eta')
   e^{-i\Omega(k_i, \eta)+i\Omega(k_i, \eta')}
   + |\beta_{\vk_i}|^2 \gamma^*(k_i, \eta)\gamma(k_i, \eta')
   e^{i\Omega(k_i, \eta)-i\Omega(k_i, \eta')} \\
   \qquad + \alpha_{\vk_i}^*\beta_{\vk_i} \gamma^*(k_i, \eta)\gamma^*(k_i, \eta')
   e^{i\Omega(k_i, \eta)+i\Omega(k_i, \eta')}
   + \alpha_{\vk_i}\beta_{\vk_i}^* \gamma(k_i, \eta)\gamma(k_i, \eta')
   e^{-i\Omega(k_i, \eta)-i\Omega(k_i, \eta')}
   \biggr)\biggr] \\
   \qquad + \text{permutations} \, . 
 \end{multline}
The largest contribution in powers of $\beta_\vk$ is of the form
 \begin{multline} \label{leadinginteractionenergyback}
  \sim\sum_i\text{Re}\biggl[
  \biggl({\phi_{,t} \ov H}\biggr)^2 {H^2 \ov M_{_\text{Planck}}^4 a(\eta)^4}
  \int {d^3k_1 \ov (2\pi)^{{3 \ov 2}}}\int {d^3k_2 \ov (2\pi)^{{3 \ov 2}}}
  \int_{\eta_{\text{{\rm II}}}(k)}^{\eta_{\riii III}(k)}d\eta'
   \text{{\small $\sum\limits_{h,j =1}^3$}}{k_1\,k_2\,k_3 \ov k_{h}^2 k_{j}^2}
   \gamma^*(k_i, \eta)\gamma^*(k_i, \eta')
    \\
   \prod_{h \neq i} \gamma(k_h, \eta)\gamma^*(k_h, \eta')
   |\alpha_{\vk_h}|^2\alpha_{\vk_i}^*\beta_{\vk_i}
   e^{2i\Omega(k_i, \eta)-i\Omega(k_i, \eta, \eta')-i\Omega(k_h, \eta, \eta')}
   \biggr] \, .
 \end{multline}
where we have defined 
 \beq
  \Omega(k_i, \eta, \eta') \equiv \Omega(k_i, \eta) - \Omega(k_i, \eta') \, .
 \eeq
The important point here is that at early times the oscillations due
to the term $e^{2i\Omega(k_i, \eta, \eta_{\riii III})}$ in equation
(\ref{leadinginteractionenergyback}) 
severely damps this contribution, and at more recent times this
already small energy density is further redshifted by the
$a(\eta)^{-4}$ factor. The same mechanism was also operating in the
modified vacuum case discussed in \cite{Holman:2007na}. 

We therefore conclude that the leading contribution from the
interaction is in fact again given by the term proportional to
$|\beta_\vk|^2$. By considering this term, changing the momentum
integration variables as
$\vk_i \to \vec y_i = {\vk_i \eta H \ov \Lambda}$ and performing the
relevant integral, we obtain the estimate 
 \beq \label{interactionsnobackreaction}
  \langle \hat \rho^{(3)} \rangle \approx \biggl({\phi_{,t} \ov H}\biggr)^2
          {H^2 \ov M_{_\text{Planck}}^4} \Delta^2 \Lambda^4 =
   \varepsilon {H^2 \ov M_{_\text{Planck}}^2} \Delta^2 \Lambda^4 \, ,
 \eeq
which does not
strengthens the constraints coming from the free contribution. 

We also have to check the backreaction coming from the higher
derivative interaction term. As before, the energy component of the
energy-momentum tensor in the interaction picture is
the Hamiltonian density, in this case of equation (\ref{hamilthigder}). For
subhorizon scales, the terms $\varphi'$ and $\partial_i\varphi$ scale
in the same way, so the contribution to the energy density is
 \begin{multline} 
  \langle \hat \rho^{(3)} \rangle  \approx 
  \text{Re} 
  \biggl[(2\pi)^3\delta(\sum_i \vk_i)
  {c^2 \phi_{,t}^{\;\;2} \ov  4 \Lambda^8 a(\eta)^6}
  \int {d^3k_1 \ov (2\pi)^{{3 \ov 2}}}\int {d^3k_2 \ov (2\pi)^{{3 \ov 2}}}
  \int_{\eta'_{\text{{\rm II}}}}^{\eta'_{\riii III}}d\eta'
   \prod_{i=1}^3 \\
   \qquad k_i\,\biggl(|\alpha_{\vk_i}|^2 \gamma(k_i, \eta)\gamma^*(k_i, \eta')
   e^{-i\Omega(k_i, \eta)+i\Omega(k_i, \eta')}
   + |\beta_{\vk_i}|^2 \gamma^*(k_i, \eta)\gamma(k_i, \eta')
   e^{i\Omega(k_i, \eta)-i\Omega(k_i, \eta')} \\
   \qquad + \alpha_{\vk_i}^*\beta_{\vk_i} \gamma^*(k_i, \eta)\gamma^*(k_i, \eta')
   e^{i\Omega(k_i, \eta)+i\Omega(k_i, \eta')}
   + \alpha_{\vk_i}\beta_{\vk_i}^* \gamma(k_i, \eta)\gamma(k_i, \eta')
   e^{-i\Omega(k_i, \eta)-i\Omega(k_i, \eta')}
   \biggr)\biggr] \\
   \qquad + \text{permutations} \, . 
 \end{multline}
By changing again variables as  
$\vk_i \to \vec y_i = {\vk_i \eta H \ov \Lambda}$ and performing the
integration, we obtain the estimate 
 \beq \label{interactionsnobackreactionhigder}
  \langle \hat \rho^{(3)} \rangle \approx \phi_{, t}^{\;\;2} \Delta^2 =
   \varepsilon {H^2 M_{_\text{Planck}}^2} \Delta^2 \, ,
 \eeq
Observe however that already for $\Lambda$ at the level of the
supersymmetric GUT scale, 
$\Lambda^4 > \varepsilon {H^2 M_{_\text{Planck}}^2}$, 
and therefore, once again, the constraints coming from the
free contribution are not strengthened.
The
final result from equations 
(\ref{energynobackreaction}), (\ref{slowrollnobackreaction}),
(\ref{interactionsnobackreaction}) is  
 \beq \label{betaconstraintnobackreaction}
  |\beta_{\vk}| \leq \sqrt{\epsilon |\mu|} {H M_{_\text{Planck}} \ov \Lambda^2}.
 \eeq

\section{Final results and conclusion}\label{conclusions}

Lorentzian
symmetry could be broken at very high energies, for example by quantum
gravity effects. The consequent
modifications in the dispersion relations could leave detectable
signatures in cosmological 
observables such as the temperature fluctuations of the CMBR, possibly
allowing an experimental investigation of these aspects of the theory.

In this article we have undergone a full general analysis of the effects of
modified dispersion relations on the bispectrum, which is the leading
contribution to the non-Gaussianities of the temperature fluctuations
in the CMBR. We have in particular focused on dispersion relations
that violate the adiabatic conditions at early times for a short
period of time. The fact that the adiabatic condition is satisfied at
the earliest times allowed us to fix unambiguously the initial conditions
(vacuum).  

The field equation in the presence of modified dispersion relations is
difficult to solve, and therefore we have been using the WKB
approximation for that scope, where possible. This approximation scheme is reliable
and has been shown to be in quantitative, beside
qualitative, agreement with the exact solutions, where available, see also
\cite{Ashoorioon:2011eg}. It also allows to obtain general results.

The universal features of the bispectrum for the modified dispersion
relations have been discussed both in
quantitative and qualitative terms. In particular, it has been shown: first,
that when there is no WKB violation at early times, the result does
not strongly differ from the standard slow-roll suppressed
one. Second, that in the case of violation of the WKB condition at early times,
the leading corrections to the bispectrum could be enhanced. The
magnitude of the enhancement factors and the 
configurations for which these appear
depend on the specific form of the dispersion relation. 

The largest enhancements for a given momenta configuration $\{k\}$ arise if there exist a
solution $y_*$ to the system of
equations~\footnote{Smaller enhancements are also possible for
  nearly critical boundary points, see footnote \ref{Nearcriticalpoints}.}
 \beq
 \partial_y^n S_{0\beta}(\{k\}, y)|_{y=y_*} = S_{0\beta}^{(n)}(\{k\}, y_*) \neq 0, \qquad 
 \partial_y^m S_{0\beta}(\{k\}, y)|_{y=y_*} = 0 \; \qquad \forall\, m < n \, ,
 \eeq
where the function $S$ is defined in terms of the comoving frequencies
$\omega$ as 
 \beq \label{phaseintegcorrect}
  S_{0 \beta}(k_1, k_2, k_3, y) = 
  \int^{y}  dy'
  \biggl(\sum_{h \neq j}\omega(k_h\,k_{_\text{m}}^{-1} , y)-\omega(k_j\,k_{_\text{m}}^{-1}, y)\biggr) \, ,
  \quad h, j \in \{1, 2, 3\}, \, \, 
  k_{_\text{m}}\!\!=\!\!\text{max}(k_{1,2,3}).
 \eeq
The schematic formula for the leading enhancements, taking care of the
constraints (\ref{betaconstraintnobackreaction}) from backreaction, is
\begin{itemize}
 \item for cubic interaction as in the Hamiltonian (\ref{HIcubic}), 
   see equations (\ref{enhancementfactors}),
 \beq 
 {\hat F_{3, \beta} \ov \hat F_{\text{3, BD}}}
  \simeq \biggl({\Lambda \ov H}\biggr)^{1-{1 \ov n}}
   {1 \ov |S_{0\beta}^{(n)}(\{k\}, y_*)|^{{1\ov n}}}
    \sqrt{\epsilon |\mu|} {H M_{_\text{Planck}} \ov \Lambda^2} \, ,
 \eeq
 \item for higher-derivative interactions as in the Hamiltonian (\ref{hamilthigder}), 
   see equations (\ref{enhancementfactorshigder}),
 \beq  \label{enhancementfactorshigderbackconstr} 
 {\hat F_{\partial,\, \beta} \ov \hat F_{\text{$\partial$, BD}}}
  \simeq \biggl({\Lambda \ov H}\biggr)^{3-{1 \ov n}}
   {y_*^2 \ov |S_{0\beta}^{(n)}(\{k\}, y_*)|^{{1\ov n}}}
    \sqrt{\epsilon |\mu|} {H M_{_\text{Planck}} \ov \Lambda^2} \, ,
 \eeq
\end{itemize}

We find differences when comparing our general results with those
found by \cite{Holman:2007na} using the modified vacuum 
approach to model transplanckian physics, which is based on imposing a
cutoff on the momenta at a certain energy scale. The
enhancements in the case of \cite{Holman:2007na} arise only for the so
called enfolded configurations, where the sum of the moduli of the
external momenta vanishes, while for all other configurations the
contribution to the bispectrum
is the standard slow-roll suppressed one. Instead, in the case of modified
dispersion relations violating WKB at early times, the
enhancements could be enjoyed by different triangle configurations,
depending on the particular dispersion relation at hand, leading
to enhanced oscillations over different areas of the triangle
space. In the lucky case, this could increase the detectability of the
signal and 
possibly somehow alleviate the suppression due to the projection onto
the two-dimensional surface following the decomposition of the
bispectrum in spherical harmonics. We leave this point for
future research.

Also, the magnitude of the enhancements does vary with 
the form of dispersion relation. In the case of cubic coupling
given by equation (\ref{HIcubic}), obtaining the same magnitude as
found in \cite{Holman:2007na} would require quite particular
conditions (that is, 
all the derivatives of the function in equation
(\ref{phaseintegcorrect}) to be zero on the 
enhanced configurations). On the other hand, for 
higher-derivative interactions as in the Hamiltonian
(\ref{hamilthigder}), the 
magnitude of the enhancements can
be larger than the one found with the modified vacuum approach, if
 \beq \label{magnmoddisplargermodvac}
  \biggl({\Lambda \ov H}\biggr)^{1-{1 \ov n}} y_*^2 > 1 \, .
 \eeq
For instance, for $\Lambda \sim 10^{3}\, H, \epsilon \sim |\mu| \sim
10^{-2}, M_{_\text{Planck}} \sim 10^{5}\, H$, $y_*^2 \sim 0.5$, $n=2$ we obtain 
${\hat F_{\partial\beta} \ov \hat F_{\text{BD}}} \sim 10^4$.

However, we stress that the presence of stationary points, in particular
satisfying the condition (\ref{magnmoddisplargermodvac}), strictly
depends on the form of the modified dispersion relation, and therefore
could not be an easily occurring feature. Nonetheless, also in the worst
case ($n = 1$ in (\ref{enhancementfactorshigderbackconstr}): no
stationary point), we can
easily have enhancements because of boundary behaviour. For instance,
in the condition of the  
example above they would be of the order
$10^3$, which is of the same magnitude as those found in the
modified vacuum framework for transplanckian physics.

\section*{Acknowledgments}

The author is thankful to Ulf Danielsson for suggesting to explore the
effects of modified dispersion relations in the bispectrum.
The author is supported by a Postdoctoral F.R.S.-F.N.R.S. research
fellowship via the Ulysses Incentive Grant for the Mobility in Science
(promoter at the Universit\'e de Mons: Per Sundell).

\appendix

\section{Appendices}

\subsection{Lorentz breaking action for the inflaton
  sector}\label{lorbreackscalact}

An action for the Lorentz breaking inflationary sector of our models
can be written as follows:
\beq \label{actionbreakingLorentz} 
  S_{\phi}=\int d^4x\sqrt{-g}(\mathcal{L}_{\phi}+
   \mathcal{L}_u), 
 \eeq
 \begin{gather}
  \mathcal{L}_{\phi}=-{1 \ov 2}g^{\mu\nu}\partial_\mu
  \phi\partial_\nu\phi - \sum_{n,p\leq n} c_{np}
  D^{2n}\phi \, D^{2p}\phi, \\
  \mathcal{L}_u=-\lambda(g^{\mu\nu}u_\mu u_\nu +1)-
  d_1F^{\mu \nu}F_{\mu \nu} 
  \quad
  F_{\mu \nu} = \nabla_\mu \,u_{\nu}-\nabla_\nu\,u_\mu \, ,  
 \end{gather}
where $u_\mu$ is the dynamical vector field necessary to properly
define the Lorentz-breaking higher-derivative terms for the inflaton
field $\phi$. 
Those terms are defined using the covariant derivative $\mathcal{D}_\mu$ associated with
what corresponds to the spatial metric seen by an observer comoving
with $u_\mu$, such that
 \beq 
 \mathcal{D}_\alpha(g_{\mu\nu}+u_\mu u_\nu) \equiv \mathcal{D}_\alpha q_{\mu\nu}= 0 \, ,
 \quad
  \mathcal{D}^{2n} \equiv (q_{\mu}^{\; \nu}\nabla_{\nu}q^{\mu\sigma}\nabla_{\sigma})^{n} \, .
 \eeq
In the action above, we have also constrained the vector
field $u_\mu$ to have unit norm using the Lagrangian multiplier
$\lambda$, as it is usually done. 

If we choose a foliation of spacetime such that $u_{\mu}$ is
orthogonal to the hypersurfaces of constant $t$, the
modified dispersion relation obtained from the equation of motion
following from the action (\ref{actionbreakingLorentz}) is
 \beq
  \omega^2_{_\text{phys}}(p)=p_{_{\rm phys}}^2+2\sum _{n,p}(-1)^{(n+p)}c_{np}
   \, p_{_\text{phys}}^{2(n+p)} \, .
\eeq

\subsection{A detailed example}\label{example}

We consider now an explicit example of dispersion relation that violates the
WKB condition at early times, given by
 \beq \label{examplefrequency}
  \omega^2_{_\text{phys}} =
  \begin{cases}
   p^2-{p^4 \ov \Lambda} & p < p_t \\
   c (p-p_0)^2+\omega_0^2 & p > p_t
  \end{cases}, \qquad p= {k \ov a(\eta)}
 \eeq
This dispersion relation is identical to the Corley-Jacobson one
with negative coefficient for $p < p_t$ \cite{Corley:1996ar}, but
avoids the problem related with the presence of 
imaginary frequencies in the original proposal thanks to the modified
behaviour for $p > p_t$. We will however not discuss its
phenomenological viability.  

\begin{figure}[ht!]
\centering
\includegraphics*[width=250pt, height=180pt]{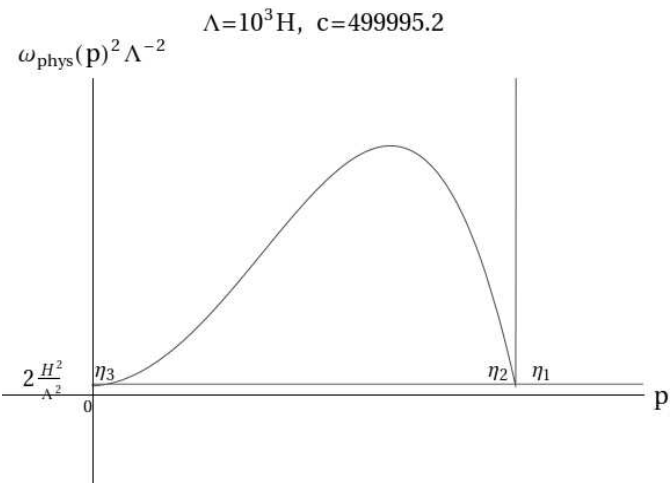}
\caption{Modified dispersion relation used for the example in section \ref{example}.}
\label{exampledispersion}
\end{figure}

The relation has a local maximum at 
$p=p_m = {\Lambda \ov \sqrt{2}}$ and is 
continuous and differentiable for
 \be \label{continuitydifferentiabilitymodquad}
  p_0 & = 2 {p_t^3 \ov c \Lambda^2}+p_t {1-c \ov c} \\
  \omega_0^2 & = p_t^2\bigl(1-{1 \ov c}\bigr)+ 
   {p_t^4 \ov \Lambda^2}\bigl({4-c \ov c}\bigr)-{4 \ov c} {p_t^6 \ov \Lambda^4}
 \label{interceptomega0modquad}
 \ee
If we assume 
$p(\eta_{_2}) \lesssim p_t$, the momenta $p(\eta_{_{2, 3}})$ of horizon
crossing (see figure \ref{exampledispersion}) are 
 \beq
  p(\eta_{_3}) = \Lambda \biggl({\sqrt{1 - \sqrt{1-8 \epsilon}} \ov \sqrt{2}}\biggr) \, ,
  \qquad
  p(\eta_{_2}) = \Lambda \biggl({\sqrt{1 + \sqrt{1-8 \epsilon}} \ov \sqrt{2}}\biggr) \, ,
  \qquad \epsilon = {H \ov \Lambda}
 \eeq
We also need to impose $0 < \omega_0^2 < \sqrt{2} H$ which yields
some conditions on $c$, $\epsilon$. We do not write
the formulas as they are complicated and of little interest here. We
make sure than in the following all the conditions are satisfied\footnote{We require
$\omega_0^2 > 0$ in order to discuss an example with the good
physical quality of not having a backreaction issue nowadays, which
would place strong bounds on $\beta_\vk$, as we have discussed in section
\ref{backreactionsec}.}. 

We now solve the equation (\ref{eqofmoUpot}) as done in section
\ref{fieldsolution} and compute the main
contribution to the bispectrum for the cubic coupling in
(\ref{HIcubic}), coming from 
the integration over the interval 
$[\eta_{\rii II}, \eta_{\riii III}]$, as discussed in section
\ref{allIII}. We neglect the other subdominant contributions.
As expected, it is found that $\eta_{\rii II (\riii III)} \simeq \eta_{_{2(3)}}$.

\begin{figure}[ht!]
\centering
\includegraphics*[width=380pt, height=380pt]{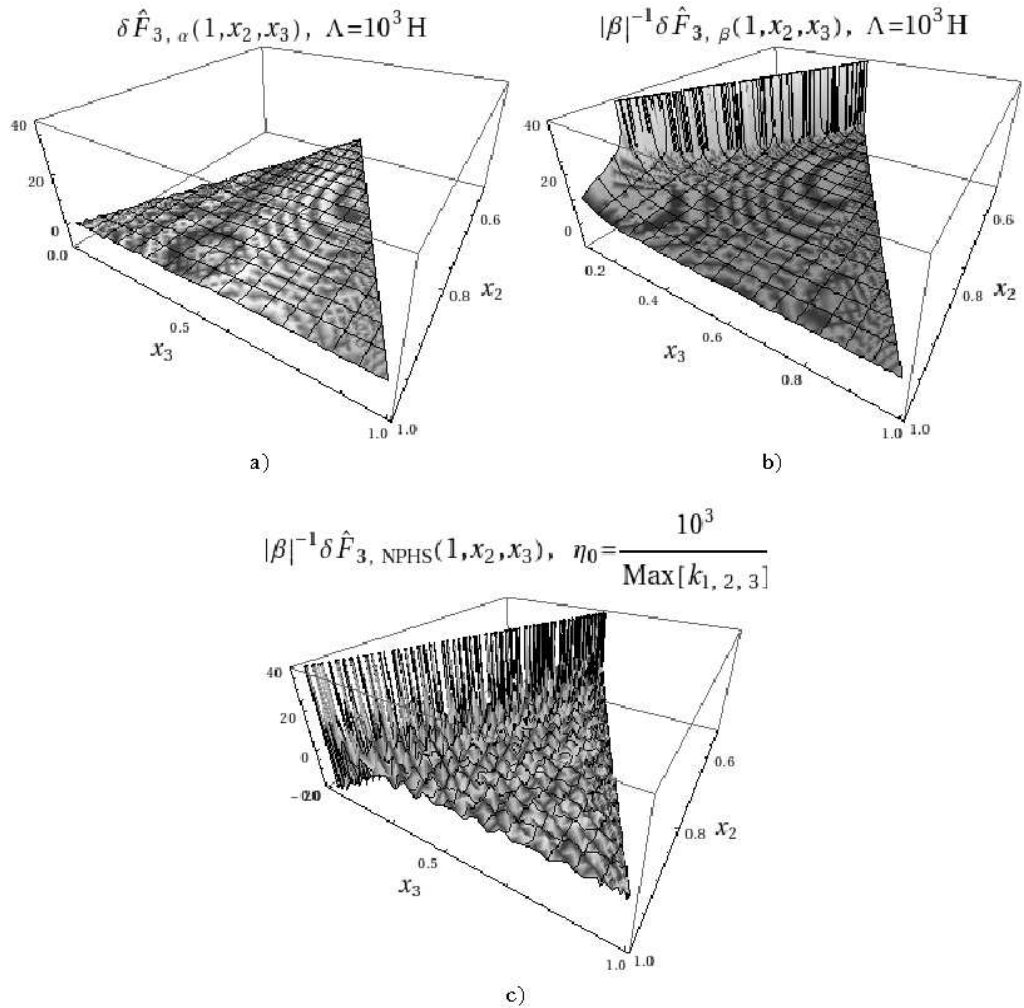}
\caption{Cubic interactions $\sim\zeta^{'3}$: a) Plot of the rescaled
  contribution to the bispectrum when all Green function 
  are in the positive-frequency branch: negligible differences with
  the standard result (tiny superimposed oscillations). b) Plot of the
  rescaled contribution to the bispectrum when one of the Green
  functions is in the 
  negative frequency branch: enhanced configurations for nearly critical
  points (see footnote
  \ref{Nearcriticalpoints}). c) Plot of the 
  rescaled contribution to the bispectrum when one of the Green
  function is in the negative frequency branch in the case of the
  modified vacuum
  modelling of transplanckian physics \cite{Holman:2007na}.}
\label{bispectrumplotsexample}
\end{figure}

We compute the contribution to the bispectrum from the cubic
interaction (\ref{HIcubic}) using the formulas 
(\ref{statcriteven}), (\ref{boundmax}) and (\ref{allthreeposIII})
adapted to our case. We do not report their final expressions
after being adapted,
since the formulas are quite long and complicated but straightforward to
be obtained. We find that in our example, the only enhanced
configurations are close to the enfolded ones, where we have nearly
critical points (see footnote \ref{Nearcriticalpoints}). We plot the
bispectrum rescaled by the standard 
slow-roll result and the modulus of $\beta_\vk$ in figure
\ref{bispectrumplotsexample} taking $\Lambda = 10^3 H$. 
The dispersion relation itself is plotted in figure \ref{exampledispersion}.  

For comparison, we
also plot the correction to the bispectrum for cubic coupling obtained
in \cite{Holman:2007na} using the modified vacuum approach to transplanckian
physics (from equation (3.17) in \cite{Holman:2007na}). We observe that the shape of the
modulations and the magnitude of 
enhancement are different between that case and the one of the
modified dispersion relation given by equation (\ref{examplefrequency}).

\begin{thebibliography}{99}

	\bibitem{Komatsu:2009kd} E.~Komatsu {\it et al.},
  arXiv:0902.4759 [astro-ph.CO]

  J.~M.~Maldacena and G.~L.~Pimentel,
  arXiv:1104.2846 [hep-th].

  I.~Antoniadis, P.~O.~Mazur and E.~Mottola,
  arXiv:1103.4164 [gr-qc].

	\bibitem{Maldacena:2002vr} J.~M.~Maldacena,
  JHEP {\bf 0305}, 013 (2003)
  [arXiv:astro-ph/0210603].

	\bibitem{Holman:2007na} R.~Holman and A.~J.~Tolley,
  JCAP {\bf 0805}, 001 (2008)
  [arXiv:0710.1302 [hep-th]].

	\bibitem{Meerburg:2009fi} P.~D.~Meerburg, J.~P.~van der Schaar and M.~G.~Jackson,
  JCAP {\bf 1002}, 001 (2010)
  [arXiv:0910.4986 [hep-th]].

	\bibitem{uncerttrans} A.~Kempf,
  Phys.\ Rev.\  D {\bf 63}, 083514 (2001)
  [arXiv:astro-ph/0009209].

  A.~Kempf and J.~C.~Niemeyer,
  Phys.\ Rev.\  D {\bf 64}, 103501 (2001)
  [arXiv:astro-ph/0103225].

  S.~F.~Hassan and M.~S.~Sloth,
  Nucl.\ Phys.\  B {\bf 674} (2003) 434
  [arXiv:hep-th/0204110].

  R.~Easther, B.~R.~Greene, W.~H.~Kinney and G.~Shiu,
  Phys.\ Rev.\  D {\bf 66}, 023518 (2002)
  [arXiv:hep-th/0204129].

  R.~Easther, B.~R.~Greene, W.~H.~Kinney and G.~Shiu,
  Phys.\ Rev.\  D {\bf 64}, 103502 (2001)
  [arXiv:hep-th/0104102].

	\bibitem{boundarytrans} K.~Schalm, G.~Shiu and J.~P.~van der Schaar,
  JHEP {\bf 0404} (2004) 076
  [arXiv:hep-th/0401164].

  K.~Schalm, G.~Shiu and J.~P.~van der Schaar,
  AIP Conf.\ Proc.\  {\bf 743} (2005) 362
  [arXiv:hep-th/0412288].

	\bibitem{momcutofftrans} U.~H.~Danielsson,
  Phys.\ Rev.\  D {\bf 66}, 023511 (2002)
  [arXiv:hep-th/0203198].

 U.~H.~Danielsson, 
 JHEP \textbf{0207} (2002) 040 [arXiv:hep-th/0205227].

	\bibitem{Seery:2005wm} D.~Seery and J.~E.~Lidsey,
  JCAP {\bf 0506} (2005) 003
  [arXiv:astro-ph/0503692].

	\bibitem{Chen:2006nt} X.~Chen, M.~x.~Huang, S.~Kachru and G.~Shiu,
  JCAP {\bf 0701} (2007) 002
  [arXiv:hep-th/0605045].

	\bibitem{Martin:2000xs} J.~Martin and R.~H.~Brandenberger,
  Phys.\ Rev.\  D {\bf 63}, 123501 (2001)
  [arXiv:hep-th/0005209].

	\bibitem{moddispreltrans} J.~Martin and R.~H.~Brandenberger,
  Phys.\ Rev.\  D {\bf 65}, 103514 (2002)
  [arXiv:hep-th/0201189].

  J.~Martin and R.~Brandenberger,
  Phys.\ Rev.\  D {\bf 68} (2003) 063513
  [arXiv:hep-th/0305161].

	\bibitem{Lemoine:2001ar} M.~Lemoine, M.~Lubo, J.~Martin and J.~P.~Uzan,
  Phys.\ Rev.\  D {\bf 65} (2002) 023510
  [arXiv:hep-th/0109128].

	\bibitem{LorentzViolReview} D.~Mattingly,
  Living Rev.\ Rel.\  {\bf 8} (2005) 5
  [arXiv:gr-qc/0502097].

  T.~Jacobson, S.~Liberati and D.~Mattingly,
  Annals Phys.\  {\bf 321} (2006) 150
  [arXiv:astro-ph/0505267].
 

	\bibitem{Jacobson:2000xp} T.~Jacobson and D.~Mattingly,
  Phys.\ Rev.\  D {\bf 64} (2001) 024028
  [arXiv:gr-qc/0007031].

	\bibitem{Mattingly:2001yd} D.~Mattingly and T.~Jacobson,
  arXiv:gr-qc/0112012.

	\bibitem{Libanov:2005yf} M.~V.~Libanov and V.~A.~Rubakov,
  JCAP {\bf 0509} (2005) 005
  [arXiv:astro-ph/0504249].

	\bibitem{Jacobson:2005bg} T.~Jacobson, S.~Liberati and D.~Mattingly,
  Annals Phys.\  {\bf 321}, 150 (2006)
  [arXiv:astro-ph/0505267].

	\bibitem{Horava:2009uw} P.~Horava,
  Phys.\ Rev.\  {\bf D79}, 084008 (2009).
  [arXiv:0901.3775 [hep-th]].

	\bibitem{lorentzviolbraneworld} D.~J.~H.~Chung and K.~Freese,
  Phys.\ Rev.\  D {\bf 61} (2000) 023511 
  [arXiv:hep-ph/9906542].

  D.~J.~H.~Chung and K.~Freese,
  Phys.\ Rev.\  D {\bf 62}, 063513 (2000)
  [arXiv:hep-ph/9910235].

  D.~J.~H.~Chung, E.~W.~Kolb and A.~Riotto,
  Phys.\ Rev.\  D {\bf 65}, 083516 (2002)
  [arXiv:hep-ph/0008126].

  C.~Csaki, J.~Erlich and C.~Grojean,
  Nucl.\ Phys.\  B {\bf 604}, 312 (2001)
  [arXiv:hep-th/0012143].

  S.~L.~Dubovsky,
  JHEP {\bf 0201}, 012 (2002)
  [arXiv:hep-th/0103205].

	\bibitem{smallsoundspeed} D.~Baumann and D.~Green,
  arXiv:1102.5343 [hep-th].

  A.~J.~Tolley and M.~Wyman,
  Phys.\ Rev.\  D {\bf 81} (2010) 043502
  [arXiv:0910.1853 [hep-th]].
 

	\bibitem{Ashoorioon:2011eg} A.~Ashoorioon, D.~Chialva and U.~Danielsson,
  arXiv:1104.2338 [hep-th]. To appear in JCAP.

	\bibitem{Corley:1996ar} S.~Corley and T.~Jacobson,
  Phys.\ Rev.\  D {\bf 54}, 1568 (1996)
  [arXiv:hep-th/9601073].

  S.~Corley,
  Phys.\ Rev.\  D {\bf 57}, 6280 (1998)
  [arXiv:hep-th/9710075].

	\bibitem{BenderOrszagMathMeth} C.~M.~Bender, S.~A.~Orszag,
  ``Asymptotic Methods for Scientists and Engeneers'', McGraw Hill, 1978,
  (Springer 1999).

	\bibitem{Martin:2002vn} J.~Martin and D.~J.~Schwarz,
  Phys.\ Rev.\  D {\bf 67} (2003) 083512
  [arXiv:astro-ph/0210090].

	\bibitem{Babich:2004gb} D.~Babich, P.~Creminelli and M.~Zaldarriaga,
  JCAP {\bf 0408}, 009 (2004)
  [arXiv:astro-ph/0405356].

	\bibitem{Erdelyi} A.~Erd\'elyi, 
  ``Asymptotic Expansions'', Dover (New York) 1956.

	\bibitem{Creminelli:2003iq} P.~Creminelli,
  JCAP {\bf 0310} (2003) 003
  [arXiv:astro-ph/0306122].
 

	\bibitem{Brandenberger:2004kx} R.~H.~Brandenberger and J.~Martin,
  Phys.\ Rev.\  D {\bf 71} (2005) 023504
  [arXiv:hep-th/0410223].

	\bibitem{Greene:2004np} B.~R.~Greene, K.~Schalm, G.~Shiu and J.~P.~van der Schaar,
  JCAP {\bf 0502} (2005) 001
  [arXiv:hep-th/0411217].

	\bibitem{Danielsson:2004xw} U.~H.~Danielsson,
  Phys.\ Rev.\  D {\bf 71} (2005) 023516
  [arXiv:hep-th/0411172].

\end{thebibliography}
\end{document}